\documentclass[zpreprint,zbstepj]{./LaTeX/zeus/zeus_paper}
\usepackage[english]{babel}

\newcommand{\ZcoosysB}{%
The ZEUS coordinate system is a right-handed Cartesian system, with the $Z$
axis pointing in the proton beam direction, referred to as the ``forward
direction'', and the $X$ axis pointing left towards the centre of HERA.
The coordinate origin is at the nominal interaction point.\xspace}

\newcommand{\ZcoosysfnB}{\footnote{\ZcoosysB}}

\newcommand{\Zctddesc}[1]{%
Charged particles are tracked in the central tracking detector (CTD)~\citeCTD,
which operates in a magnetic field of $1.43\Tesla$ provided by a thin 
superconducting solenoid. The CTD consists of 72~cylindrical drift chamber 
layers, organized in nine superlayers covering the polar-angle#1 region 
\mbox{$15^\circ<\theta<164^\circ$}. The transverse-momentum resolution for
full-length tracks is $\sigma(p_T)/p_T=0.0058p_T\oplus0.0065\oplus0.0014/p_T$,
with $p_T$ in $\Gev$.}
\newcommand{\Zcaldesc}{%
The high-resolution uranium--scintillator calorimeter (CAL)~\citeCAL consists 
of three parts: the forward (FCAL), the barrel (BCAL) and the rear (RCAL)
calorimeters. Each part is subdivided transversely into towers and
longitudinally into one electromagnetic section (EMC) and either one (in RCAL)
or two (in BCAL and FCAL) hadronic sections (HAC). The smallest subdivision of
the calorimeter is called a cell.  The CAL energy resolutions, as measured under
test-beam conditions, are $\sigma(E)/E=0.18/\sqrt{E}$ for electrons and
$\sigma(E)/E=0.35/\sqrt{E}$ for hadrons, with $E$ in $\Gev$.}

\chardef\usc=95
\chardef\til=126
\catcode`\@=11 
\DeclareRobustCommand\xdotspace{\futurelet\@let@token\@xdotspace}
\def\@xdotspace{%
  \ifx\@let@token.\else
  \ifx\@let@token\bgroup.\else
  \ifx\@let@token\egroup.\else
  \ifx\@let@token\/.\else
  \ifx\@let@token\ .\else
  \ifx\@let@token~.\else
  \ifx\@let@token!.\else
  \ifx\@let@token,.\else
  \ifx\@let@token:.\else
  \ifx\@let@token;.\else
  \ifx\@let@token?.\else
  \ifx\@let@token/.\else
  \ifx\@let@token'.\else
  \ifx\@let@token).\else
  \ifx\@let@token-.\else
  \ifx\@let@token\@xobeysp.\else
  \ifx\@let@token\space.\else
  \ifx\@let@token\@sptoken.\else
   .\space
   \fi\fi\fi\fi\fi\fi\fi\fi\fi\fi\fi\fi\fi\fi\fi\fi\fi\fi}
\catcode`\@=12 

\newcommand{\stru}[2]{%
   \relax\ifmmode\hbox{\vrule height#1 depth#2 width0pt}%
   \else\vrule height#1 depth#2 width0pt\fi}

\newcommand{\Ronum}[1]{\uppercase\expandafter{\romannumeral#1}}
\newcommand{\ronum}[1]{\expandafter{\romannumeral#1}}
\DeclareRobustCommand{\LaTeXZ}{%
  \LaTeX\kern-.05em4\kern-.1em
  {\raisebox{-0.2ex}{$\scriptstyle\text{ZEUS}$}}\xspace}

\DeclareMathAlphabet{\mathbf}{OT1}{cmr}{bx}{sl}
\newcommand{\eVdist}{\kern-0.06667em}

\newcommand{\Gev}{{\text{Ge}\eVdist\text{V\/}}}

\newcommand{\mev}{{\,\text{Me}\eVdist\text{V\/}}}
\newcommand{\gev}{{\,\text{Ge}\eVdist\text{V\/}}}

\newcommand{\pbi}{\,\text{pb}^{-1}}

\newcommand{\met}{\,\text{m}}

\newcommand{\Tesla}{\,\text{T}}

\newcommand{\slashfrac}[2]{%
  \raisebox{0.5ex}{\ensuremath #1}\kern-0.12em/\kern-0.08em
  \raisebox{-.8ex}{\ensuremath #2}}

\newcommand{\sqr}[3]{%
    {\vcenter{\hrule height.#3ex\hbox{\vrule width.#2ex height#1ex
     \kern#1ex\vrule width.#3ex}\hrule height.#2ex}}}

\catcode`\@=11 
\newcommand{\parenbar}{\mathpalette\p@renb@r}
\def\p@renb@r#1#2{\vbox{%
  \ifx#1\scriptscriptstyle \dimen@.7em\dimen@ii.2em\else
  \ifx#1\scriptstyle \dimen@.8em\dimen@ii.25em\else
  \dimen@1em\dimen@ii.4em\fi\fi \offinterlineskip
  \ialign{\hfill##\hfill\cr
    \vbox{\hrule width\dimen@ii}\cr
    \noalign{\vskip-.3ex}%
    \hbox to\dimen@{$\mathchar300\hfil\mathchar301$}\cr
    \noalign{\vskip-.3ex}%
    $#1#2$\cr}}}
\catcode`\@=12 

\newcommand{\IP}{{\rm I$\kern-0.01667em$P}\xspace}

\mathchardef\qsm=63
\mathchardef\pls=43
\mathchardef\mns=512
\mathchardef\plm=518
\mathchardef\eql=61
\mathchardef\smallleft=300
\mathchardef\smallright=301
\mathchardef\les=316
\mathchardef\gre=318
\mathchardef\leq=532
\mathchardef\grq=533

\catcode`\@=11 
\newcounter{pict@width}
\newcounter{pict@height}
\newlength{\pict@scale}
\newcommand{\psfigadd}[4]{%
\setcounter{pict@width}{1*\ratio{#2+\pict@scale/2}{\pict@scale}}
\setcounter{pict@height}{1*\ratio{#3+\pict@scale/2}{\pict@scale}}
\setlength{\unitlength}{\pict@scale}
\hbox to #2{\hspace{-\fill}\begin{picture}(\thepict@width,\thepict@height)
\put(0,0){\psfig{figure=#1,width=#2,height=#3,clip=}}
\SetScale{0.283466457}
\SetWidth{1.763889}
{#4}
\end{picture}}
}
\newcounter{pict@widthfst}
\newcounter{pict@widthscd}
\newcounter{pict@widthtot}
\newcommand{\psfigaddtwo}[7]{%
\setcounter{pict@widthfst}{1*\ratio{#2+\pict@scale/2}{\pict@scale}}
\setcounter{pict@widthscd}{1*\ratio{#2+#4+\pict@scale/2}{\pict@scale}}
\setcounter{pict@widthtot}{1*\ratio{#2+#4+#6+\pict@scale/2}{\pict@scale}}
\setcounter{pict@height}{1*\ratio{#3+\pict@scale/2}{\pict@scale}}
\setlength{\unitlength}{\pict@scale}
\hbox{\hspace{-\fill}\begin{picture}(\thepict@widthtot,\thepict@height)
\put(0,0){\psfig{figure=#1,width=#2,height=#3,clip=}}
\put(\thepict@widthscd,0){\psfig{figure=#5,width=#6,height=#3,clip=}}
\SetScale{0.283466457}
\SetWidth{1.763889}
{#7}
\end{picture}}
}
\newcommand{\psfigror}[4]{%
\setcounter{pict@width}{1*\ratio{#2+\pict@scale/2}{\pict@scale}}
\setcounter{pict@height}{1*\ratio{#3+\pict@scale/2}{\pict@scale}}
\setlength{\unitlength}{\pict@scale}
\hbox{\begin{picture}(\thepict@width,\thepict@height)
\put(0,\thepict@height){\psfig{figure=#1,width=#3,height=#2,clip=,angle=270}}
\SetScale{0.283466457}
\SetWidth{1.763889}
{#4}
\end{picture}}
}
\newcommand{\psfigrol}[4]{%
\setcounter{pict@width}{1*\ratio{#2+\pict@scale/2}{\pict@scale}}
\setcounter{pict@height}{1*\ratio{#3+\pict@scale/2}{\pict@scale}}
\setlength{\unitlength}{\pict@scale}
\hbox{\begin{picture}(\thepict@width,\thepict@height)
\put(0,0){\psfig{figure=#1,width=#3,height=#2,clip=,angle=90}}
\SetScale{0.283466457}
\SetWidth{1.763889}
{#4}
\end{picture}}
}
\catcode`\@=12 
\newlength\listtextwidth

\catcode`\@=11 
\newlength{\@tabfninsert}
\newlength{\@tabfnwidth}
\newcommand{\tabfootnote}[2]{%
  \setlength{\@tabfninsert}{0.8em}
  \setlength{\@tabfnwidth}{\textwidth}
  \addtolength{\@tabfnwidth}{-\@tabfninsert}
  \addtolength{\@tabfnwidth}{-0.4em}
  \noindent\makebox[\@tabfninsert][r]{\footnotesize$^{#1}$\hfil}\hfill%
  \parbox[t]{\@tabfnwidth}{\footnotesize #2\hfill}}
\catcode`\@=12 

\newcommand{\dsp}        {\mbox{$D^{\ast +}$}}
\newcommand{\dspm}       {\mbox{$D^{\ast \pm}$}}

\newcommand{\dssp}       {\mbox{$D_s^+$}}

\newcommand{\dsspm}      {\mbox{$D_s^{\pm}$}}
\newcommand{\dz}         {\mbox{$D^{0}$}}
\newcommand{\dc}         {\mbox{$D^+$}}
\newcommand{\dcpm}       {\mbox{$D^{\pm}$}}

\newcommand{\lcp}        {\mbox{$\Lambda_c^+$}}
\newcommand{\lcpm}       {\mbox{$\Lambda_c^\pm$}}
\newcommand{\fcdz}       {\mbox{$f(c \rightarrow D^0)$}}
\newcommand{\fcdc}       {\mbox{$f(c \rightarrow D^+)$}}
\newcommand{\fcdss}      {\mbox{$f(c \rightarrow D_s^+)$}}
\newcommand{\fclc}       {\mbox{$f(c \rightarrow \Lambda_c^+)$}}
\newcommand{\fcds}       {\mbox{$f(c \rightarrow D^{\ast +})$}}
\newcommand{\br}         {\mbox{${\cal B}_{D^{\ast +}\rightarrow D^0 \pi^+}$}}
\def\dsk3pi{ {\dsp}~\rightarrow~\dz~\pi^{+}_{s}%
        \rightarrow~(K^{-}~\pi^{+}~\pi^{+}~\pi^{-})~\pi^{+}_{s} }
\def\et10t{ E_T^{\theta > 10^\circ}}
\def\etw10{ E_T^{\theta > 10}}
\newcommand{\lum}{\mbox{$\cal L$}}
\newcommand{\bran}{\mbox{$\cal B$}}
\newcommand{\acc}{\mbox{$\cal A$}}
\def\citeCTD{{\cite{%
nim:a279:290,*npps:b32:181,*nim:a338:254%
}}\xspace}
\def\citeCAL{{\cite{%
nim:a309:77,*nim:a309:101,*nim:a321:356,*nim:a336:23%
}}\xspace}

\begin{document}
\prepnum{{DESY--05--147}}

\title{
Measurement of charm fragmentation ratios and fractions in
photoproduction at HERA
}                                                       
                    
\author{ZEUS Collaboration}

\date{August 2005}

\abstract{
The production of $\dsp$, $\dz$, $\dc$, $\dssp$ and $\lcp$
charm hadrons and their antiparticles in
$e p$ scattering
at HERA was measured with the ZEUS detector using an 
integrated luminosity of $79\pbi$. The measurement has been performed 
in the photoproduction regime
with the exchanged-photon virtuality $Q^2 < 1\gev^2$ and
for photon-proton centre-of-mass energies in the range $130 < W < 300\gev$.
The charm hadrons were 
reconstructed in the range
of transverse momentum $p_T(D,\Lambda_c) > 3.8\gev$
and pseudorapidity $|\eta(D,\Lambda_c)| < 1.6$.
The production cross sections were
used to determine the ratio of neutral and charged $D$-meson 
production rates, $R_{u/d}$, the strangeness-suppression factor, $\gamma_s$, 
and the fraction of charged $D$ mesons produced in a vector state,
$P^d_{\rm v}$.
The measured $R_{u/d}$ and $\gamma_s$ values agree
with those obtained in deep inelastic scattering and in $e^+e^-$ annihilations.
The measured $P^d_{\rm v}$ value is smaller than, but consistent with,
the previous measurements.
The fractions of $c$ quarks hadronising as a particular charm hadron, 
$f(c \rightarrow D, \Lambda_c)$, were derived
in the given kinematic range.
The measured open-charm fragmentation fractions are consistent
with previous results,
although the measured $f(c\rightarrow D^{*+})$ is smaller
and $f(c\rightarrow \Lambda_c^+)$ is larger than those
obtained in  $e^+e^-$ annihilations.
These results generally support the hypothesis that fragmentation
proceeds independently of the hard sub-process.
}

\makezeustitle

\def\3{\ss}                                                                                        
\pagenumbering{Roman}                                                                              
                                                   %
\begin{center}                                                                                     
{                      \Large  The ZEUS Collaboration              }                               
\end{center}                                                                                       
  S.~Chekanov,                                                                                     
  M.~Derrick,                                                                                      
  S.~Magill,                                                                                       
  S.~Miglioranzi$^{   1}$,                                                                         
  B.~Musgrave,                                                                                     
  \mbox{J.~Repond},                                                                                
  R.~Yoshida\\                                                                                     
 {\it Argonne National Laboratory, Argonne, Illinois 60439-4815}, USA~$^{n}$                       
\par \filbreak                                                                                     
  M.C.K.~Mattingly \\                                                                              
 {\it Andrews University, Berrien Springs, Michigan 49104-0380}, USA                               
\par \filbreak                                                                                     
  N.~Pavel, A.G.~Yag\"ues Molina \\                                                                
  {\it Institut f\"ur Physik der Humboldt-Universit\"at zu Berlin,                                 
           Berlin, Germany}
% ...
\newpage                                                                                           
$^{\    1}$ also affiliated with University College London, UK \\                                  
% ...
                                                           %
                                                           %
\newpage   
                                                           %
                                                           %
\begin{tabular}[h]{rp{14cm}}                                                                       
$^{a}$ &  supported by the Natural Sciences and Engineering Research Council of Canada (NSERC) \\  
$^{b}$ &  supported by the German Federal Ministry for Education and Research (BMBF), under        
          contract numbers HZ1GUA 2, HZ1GUB 0, HZ1PDA 5, HZ1VFA 5\\                                
$^{c}$ &  supported in part by the MINERVA Gesellschaft f\"ur Forschung GmbH, the Israel Science   
          Foundation (grant no. 293/02-11.2), the U.S.-Israel Binational Science Foundation and    
          the Benozyio Center for High Energy Physics\\                                            
$^{d}$ &  supported by the German-Israeli Foundation and the Israel Science Foundation\\           
$^{e}$ &  supported by the Italian National Institute for Nuclear Physics (INFN) \\                
$^{f}$ &  supported by the Japanese Ministry of Education, Culture, Sports, Science and Technology 
          (MEXT) and its grants for Scientific Research\\                                          
$^{g}$ &  supported by the Korean Ministry of Education and Korea Science and Engineering          
          Foundation\\                                                                             
$^{h}$ &  supported by the Netherlands Foundation for Research on Matter (FOM)\\                   
$^{i}$ &  supported by the Polish State Committee for Scientific Research, grant no.               
          620/E-77/SPB/DESY/P-03/DZ 117/2003-2005 and grant no. 1P03B07427/2004-2006\\             
$^{j}$ &  partially supported by the German Federal Ministry for Education and Research (BMBF)\\   
$^{k}$ &  supported by RF Presidential grant N 1685.2003.2 for the leading scientific schools and  
          by the Russian Ministry of Education and Science through its grant for Scientific        
          Research on High Energy Physics\\                                                        
$^{l}$ &  supported by the Spanish Ministry of Education and Science through funds provided by     
          CICYT\\                                                                                  
$^{m}$ &  supported by the Particle Physics and Astronomy Research Council, UK\\                   
$^{n}$ &  supported by the US Department of Energy\\                                               
$^{o}$ &  supported by the US National Science Foundation\\                                        
$^{p}$ &  supported by the Polish Ministry of Scientific Research and Information Technology,      
          grant no. 112/E-356/SPUB/DESY/P-03/DZ 116/2003-2005 and 1 P03B 065 27\\                  
$^{q}$ &  supported by FNRS and its associated funds (IISN and FRIA) and by an Inter-University    
          Attraction Poles Programme subsidised by the Belgian Federal Science Policy Office\\     
\end{tabular}                                                                                      
                                                           %
                                                           %

\pagenumbering{arabic} 
\pagestyle{plain}
\section{Introduction}
\label{sec-int}

Charm quark production has been extensively studied at HERA
using $\dspm$ and $\dsspm$
mesons~\cite{epj:c6:67,np:b545:21,epj:c12:35,pl:b481:213,pl:b528:199}.
The data have been compared with theoretical predictions
by assuming the universality of charm fragmentation and using
the charm fragmentation characteristics
obtained in $e^+e^-$ annihilation
for the calculations of charm production in $ep$ scattering.
However, the charm production mechanisms are not the same
in different collisions. In particular, $c{\bar c}$ pairs
in $e^+e^-$ annihilation are produced dominantly
in a colour-singlet state, which is
not the case for $ep$ scattering. Thus, it is important to test
the charm-fragmentation universality by measuring the charm
fragmentation characteristics at HERA.

In this paper, the measurement of the production
of the weakly decaying charm ground states, the $\dz$, $\dc$,
$\dssp$ pseudo-scalar mesons and the $\lcp$ baryon, is presented.
The production of the charm vector meson $\dsp$ has also been studied. 
The antiparticles of these charm hadrons have been measured as
well\footnote{Hereafter, charge conjugation is implied.}.
The measurement has been performed in
$e p$ scattering at HERA
in the photoproduction regime
with exchanged-photon virtuality, $Q^2$, close to zero and for
photon-proton centre-of-mass energies in the range $130<W<300\,$GeV.
The measured production cross sections have 
been used to determine the ratio of neutral and charged $D$ meson 
production rates, $R_{u/d}$, the strangeness-suppression factor, $\gamma_s$, 
and the fraction of charged $D$ mesons produced in
a vector state, $P^d_{\rm v}$.
The fractions of $c$ quarks hadronising as a particular charm hadron, 
$f(c \rightarrow D, \Lambda_c)$, have been calculated
in the accepted kinematic range.
The open-charm fragmentation fractions in photoproduction
are reported here for the first time.
The results have been compared with the previous HERA
measurements
of the charm fragmentation characteristics in
photoproduction~\cite{pl:b481:213}
and in deep inelastic scattering (DIS)
with $Q^2>2\,$GeV$^2$~\cite{epj:c38:447}.
To compare the results with those obtained in charm production
in $e^+e^-$
annihilations,
the $f(c \rightarrow D, \Lambda_c)$ fractions compiled
previously~\cite{hep-ex-9912064p} have been updated using recent
values~\cite{pl:b592:1} of the relevant branching ratios.

\section{Experimental set-up}
\label{sec:expset}

The analysis was performed with data taken
by the ZEUS Collaboration
from 1998 to 2000.
In this period, HERA collided electrons or positrons\footnote{
From now on, the word ``electron'' is used as a generic term
for electrons and positrons.}
with energy $E_e=27.5\gev$ and protons with energy $E_p=920\gev$.
The results are based on a sum of the $e^-p$ and $e^+p$ samples corresponding
to a total integrated luminosity of
$78.6\pm1.7\pbi$.
Due to trigger considerations,
$\dc$ and $\lcp$ production was measured using only
the $e^+p$ sample
corresponding
to an integrated luminosity of
$65.1\pm1.5\pbi$.

A detailed description of the ZEUS detector can be found 
elsewhere~\cite{zeus:1993:bluebook}. A brief outline of the 
components most relevant to this analysis is given
below.

\Zctddesc\ZcoosysfnB
~To estimate the energy loss per unit length, $dE/dx$, of particles in
the CTD~\cite{pl:b481:213,epj:c18:625},
the truncated mean of the anode-wire pulse heights was calculated,
which
removes the lowest $10\%$ and at least the highest $30\%$
depending on the number of saturated hits.
The measured $dE/dx$ values were normalised to
the $dE/dx$ peak position for tracks
with momenta $0.3<p<0.4\,$GeV,
the region of minimum ionisation for pions.
Henceforth $dE/dx$ is quoted in units of minimum
ionising particles (mips).
The resolution of the $dE/dx$ measurement
for full-length tracks is about $9\%$.

\Zcaldesc

The luminosity was determined from the rate of the bremsstrahlung process
$ep \rightarrow e \gamma p$, where the photon was measured with a 
lead--scintillator calorimeter~\cite{desy-92-066,*zfp:c63:391,*acpp:b32:2025} 
located at $Z = -107\met$.

\section{Event simulation}
\label{sec-simul}

Monte Carlo (MC) samples of charm and beauty events
were produced with
the {\sc Pythia} 6.156~\cite{cpc:82:74},
{\sc Rapgap}~2.0818~\cite{cpc:86:147}
and {\sc Herwig}~6.301~\cite{cpc:67:465,*jhep:0101:010}
event generators.
The generation, based on leading-order matrix elements,
includes direct photon processes,
in which the photon couples
as a point-like object in the hard scatter,
and resolved photon processes, where the photon acts as a source
of partons, one of which participates in the hard scattering process.
Initial- and final-state parton showering is added to simulate
higher-order processes.
The CTEQ5L~\cite{epj:c12:375} and GRV~LO~\cite{pr:d46:1973} parametrisations
were used for the proton and photon structure functions, respectively.
The charm and bottom quark masses were set to $1.5\,$GeV and $4.75\,$GeV,
respectively.
Events for all processes were generated in proportion to the predicted
MC cross sections.
The Lund string model~\cite{prep:97:31}
as implemented in {\sc Jetset}~\cite{cpc:82:74}
was used for hadronisation in {\sc Pythia} and {\sc Rapgap}.
The Bowler modification~\cite{zfp:c11:169}
of the Lund symmetric fragmentation function~\cite{zfp:c20:317}
was used for the charm and bottom quark fragmentation.
In {\sc Herwig},
the cluster model~\cite{np:b238:492}
was used for hadronisation.
The fraction of charged $D$ mesons produced in
a vector state was set to $0.6$ for all MC samples.

The {\sc Pythia} and {\sc Rapgap} generators
were tuned to describe the photoproduction and DIS regimes,
respectively.
Consequently,
the {\sc Pythia} events, generated with $Q^2<0.6\gev^2$,
were combined with the {\sc Rapgap} events, generated with $Q^2>0.6\gev^2$.
Diffractive events, characterised by
a large rapidity gap
between the proton at high rapidities and the centrally-produced
hadronic system,
were generated using the {\sc Rapgap} generator in the diffractive mode
and combined with the non-diffractive MC sample.
The contribution of diffractive events was estimated by fitting
the $\eta_{\rm max}$ distribution\footnote{
The quantity $\eta_{\rm max}$ is defined as the pseudorapidity
of the CAL energy deposit with the lowest polar angle and an energy
above $400\mev$.}
of the data with a linear
combination of the non-diffractive and diffractive MC samples.
The combined sample was used to evaluate the nominal acceptances.
The {\sc Herwig} MC sample,
generated over the full range of $Q^2$ values,
was used to estimate the model dependence
of the acceptance corrections.

To ensure a good description of the data,
the transverse momenta, $p_T(D,\Lambda_c)$, and pseudorapidity,
$\eta(D,\Lambda_c)$, distributions were reweighted
for both combined {\sc Pythia}+{\sc Rapgap} and {\sc Herwig} MC samples.
The reweighting
factors were tuned using a large $D^{*\pm}$ sample~\cite{epj:c38:29}.
The effect of the reweighting on the measured fragmentation ratios and
fractions was small; the reweighting uncertainty was included
when estimating the model dependence of the acceptance corrections. 

The generated events were passed through a full simulation
of the detector using {\sc Geant} 3.13~\cite{tech:cern-dd-ee-84-1}
and processed with the same reconstruction program as used for the data.

\section{Event selection}
\label{sec-selec}

A three-level trigger system was used
to select events online~\cite{zeus:1993:bluebook,uproc:chep:1992:222}.
The first- and second-level trigger used CAL and CTD
data to select $ep$ collisions and to reject beam-gas events.
At the third level,
where the full event information was available,
at least one
reconstructed
charm-hadron candidate was required.
The efficiency of the online charm-hadron reconstruction,
determined relative to the efficiency of the offline reconstruction,
was above $95\%$.

Photoproduction events were selected by requiring that no scattered
electron was identified in the CAL~\cite{pl:b322:287}.
The Jacquet-Blondel~\cite{proc:epfacility:1979:391} estimator of $W$,
$W_{\rm JB} = \sqrt{2 E_{p} (E-p_{Z})}$,
was used, where $E-p_{Z}={\Sigma_{i}(E-p_{Z})_{i}}$
and the sum $i$ runs over all final state energy-flow objects~\cite{briskinu:phd:1998} produced from
charged tracks,
as measured in the CTD, and energy clusters measured in the CAL.
After correcting for detector effects, the most important of which were
energy losses in inactive material in front of the CAL and particle losses
in the beam pipe~\cite{pl:b322:287,pl:b349:225}, 
events were selected in the interval $130<W<300\,$GeV.
The lower limit was set by the trigger requirements, while the upper
limit was imposed to suppress remaining DIS events
with an unidentified scattered
electron in the CAL~\cite{pl:b322:287}.
Under these conditions, the
photon virtuality lies below $1\,$GeV$^2$.
The median $Q^{2}$ value was estimated from a
Monte Carlo simulation to be about $3 \times 10^{-4}\,$GeV$^2$.

\section{Reconstruction of charm hadrons}
\label{sec-rec}

The production of
$\dsp$, $\dz$, $\dc$, $\dssp$ and $\lcp$
charm hadrons
was measured in
the range of transverse momentum $p_T(D,\Lambda_c)>3.8\gev$
and pseudorapidity $|\eta(D,\Lambda_c)|<1.6$.
Charm hadrons were reconstructed using tracks measured
in the CTD and assigned to the reconstructed event vertex.
To ensure good momentum resolution, each track was required
to reach at least the third superlayer of the CTD.
The combinatorial background was significantly reduced by requiring
$p_T(D)/\et10t > 0.2$
and $p_T(\Lambda_c)/\et10t > 0.25$
for charm mesons and baryons, respectively.
The transverse energy was calculated as
$\et10t={\Sigma_{i,\theta_i > 10^\circ}(E_{i}\sin \theta_i})$,
where the sum runs over all energy deposits in the CAL
with the polar angle
$\theta$ above $10^\circ$.
Further background reduction was achieved by imposing cuts
on the transverse momenta and decay angles
of the charm-hadron decay products.
The cut values were tuned using
MC simulation
to enhance signal over background ratios
while keeping acceptances high.

The details of the
reconstruction of the five charm-hadron samples
are given in the next sub-sections.

\subsection{Reconstruction of $\dz$ mesons}
\label{sec-recd0}

The $\dz$ mesons were reconstructed
from the decay
$D^0 \rightarrow K^- \pi^+$.
In each event, tracks with opposite charges and
$p_T>0.8\gev$
were combined in pairs to form $\dz$ candidates.
The nominal kaon and pion masses were assumed
in turn for each track and
the pair invariant mass, $M(K \pi)$, was calculated.
The distribution of the cosine of the $\dz$ decay angle
(defined as the angle $\theta^*(K)$ between the kaon
in the $K \pi$ rest frame
and the $K \pi$ line of flight in the laboratory frame) is flat,
whereas the combinatorial background peaks
in the forward and backward directions.
To suppress the background,
$|\cos\theta^*(K)|<0.85$ was required.

For selected $\dz$ candidates, a search was performed for a track
that could be a ``soft'' pion ($\pi_s$) in a $\dsp \rightarrow \dz \pi^+_s$
decay.
The soft pion was required to have $p_T>0.2\gev$ and a charge opposite
to that of the particle taken as a kaon.
The $p_T$ cut was raised to $0.25\gev$ for a data subsample,
corresponding to an integrated luminosity of
$16.9\pm0.4\pbi$, for which the low-momentum track reconstruction
efficiency was smaller due to the operating conditions
of the CTD~\cite{nim:a515:37}.
The corresponding $\dz$ candidate was assigned to a class of candidates
``with $\Delta M$ tag'' if the mass difference,
$\Delta M=M(K \pi \pi_s)-M(K \pi)$, was in the range
$0.143<\Delta M<0.148\gev$.
All remaining $\dz$ candidates were assigned
to a class of candidates ``without $\Delta M$ tag''.
For $\dz$ candidates with $\Delta M$ tag,
the kaon and pion mass assignment was fixed
by the track-charge requirements.
For $\dz$ mesons without $\Delta M$ tag, the mass
assignment is ambiguous.
The pion and kaon masses can therefore be
assigned to two tracks either correctly, producing a signal peak,
or incorrectly, producing a wider reflected signal.
To remove this reflection,
the mass distribution, obtained for $\dz$ candidates with $\Delta M$ tag
and an opposite mass assignment to the kaon and pion tracks,
was subtracted from the $M(K \pi)$ distribution for all $\dz$ candidates
without $\Delta M$ tag. The subtracted mass distribution was normalised
to the ratio of numbers of $\dz$ mesons without and with
$\Delta M$ tag obtained from a fit described below.

\begin{figure}[hbtp]
\epsfysize=18cm
\vspace*{-1.0cm}
\centerline{\epsffile{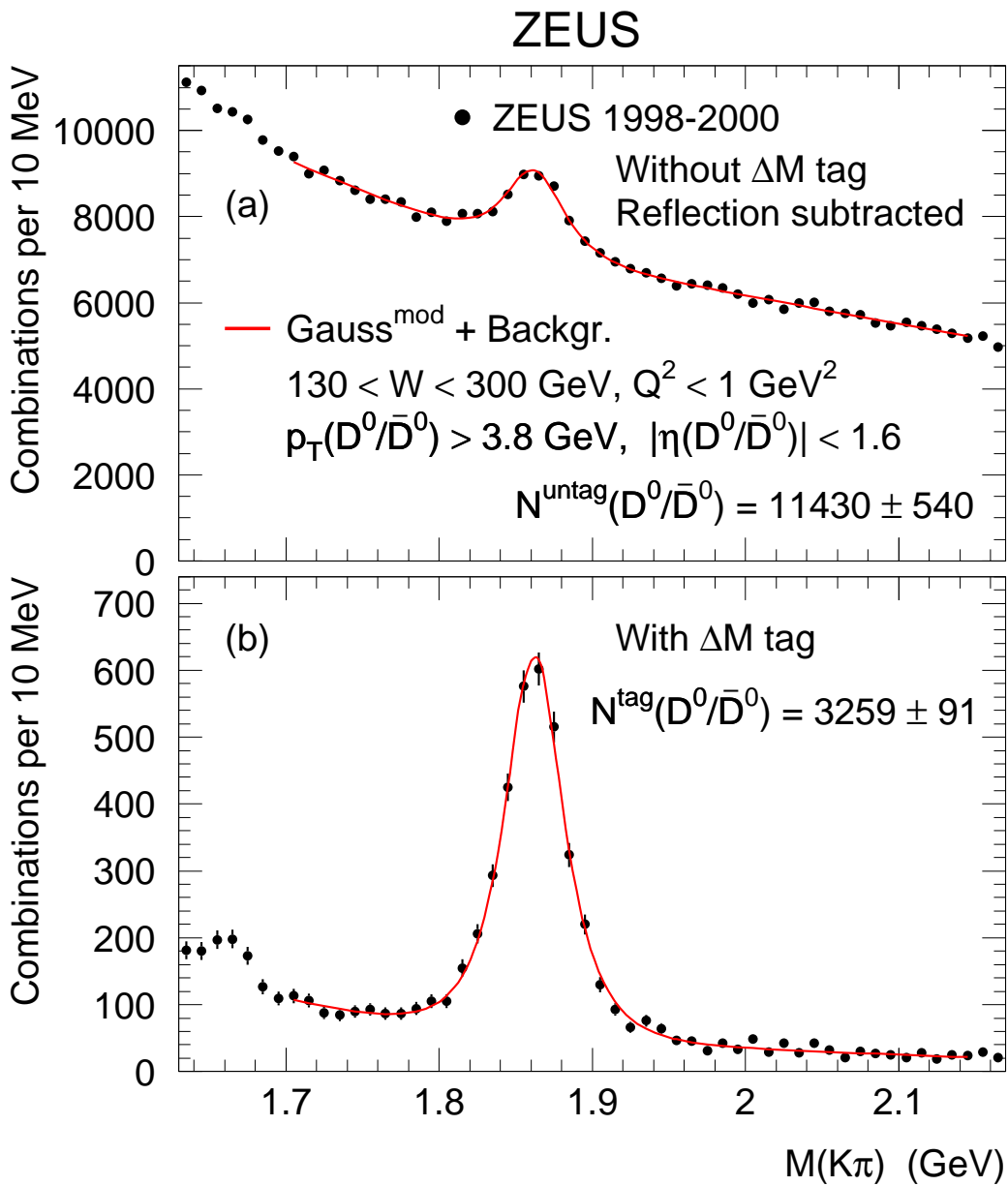}}
\caption{
The $M(K \pi)$ distributions (dots) for (a) the $D^0 / {\bar D^0}$ candidates
without $\Delta M$ tag, obtained after the reflection subtraction (see text),
and for (b) the $D^0 / {\bar D^0}$  candidates
with $\Delta M$ tag.
The solid curves represent a fit to the sum of a modified Gaussian
function and a background function.
}
\label{fig:d0}
\end{figure}

Figure~\ref{fig:d0} shows the $M(K \pi)$ distribution for $\dz$ candidates
without $\Delta M$ tag, obtained after the reflection subtraction,
and the $M(K \pi)$ distribution for $\dz$ candidates
with $\Delta M$ tag.
Clear signals are seen at the nominal value of $M(\dz)$ in both
distributions.
The distributions were fitted simultaneously assuming the same shape for
signals in both distributions. To describe the shape, a ``modified''
Gaussian function was used:
\begin{equation}
{\rm Gauss}^{\rm mod}\propto \exp [-0.5 \cdot x^{1+1/(1+0.5 \cdot x)}],
\label{eq:gausmod}
\end{equation}
where $x=|[M(K\pi)-M_0]/\sigma|$.
This functional form described both data and MC signals well.
The signal position, $M_0$,
and width, $\sigma$, as well as the numbers of $\dz$ mesons in each signal
were free parameters of the fit.
Monte Carlo studies showed that background shapes
in both distributions are compatible with being
linear in the mass range above the signals.
For smaller $M(K \pi)$ values, the background
shapes exhibit an exponential enhancement due to contributions from
other $\dz$ decay modes and other $D$ mesons.
Therefore the background shape in the fit was described by the form
$[A+B\cdot M(K \pi)]$ for $M(K \pi)>1.86\gev$ and
$[A+B\cdot M(K \pi)]\cdot \exp\{C\cdot[M(K \pi)-1.86]\}$
for $M(K \pi)<1.86\gev$.
The free parameters $A$, $B$ and $C$ were assumed to be independent for the
two $M(K \pi)$ distributions.
The numbers of $\dz$ mesons yielded by the fit were
$N^{\rm untag}(\dz)=11430\pm540$ and
$N^{\rm tag}(\dz)=3259\pm91$ for selections without and with $\Delta M$ tag,
respectively.

\subsection{Reconstruction of additional $\dsp$ mesons}
\label{sec-recds}

The $\dsp \rightarrow \dz \pi^+_s$
events
with $p_T(\dsp)>3.8\gev$ and $|\eta(\dsp)|<1.6$
can be considered as a sum of two subsamples:
events with the $\dz$ having $p_T(\dz)>3.8\gev$ and $|\eta(\dz)|<1.6$,
and events with the $\dz$ outside
of that kinematic range.
The former sample is represented by
$\dz$ mesons reconstructed with $\Delta M$ tag, as discussed in the previous
section. The latter sample of ``additional'' $\dsp$ mesons
was obtained using
the same $D^0 \rightarrow K^- \pi^+$ decay channel
and an independent selection described below.

In each event, tracks with opposite charges and $p_T>0.4\gev$
were combined in pairs to form $\dz$ candidates.
To calculate the pair invariant mass, $M(K \pi)$,
kaon and pion masses were assumed in turn for each track.
Only $\dz$ candidates which satisfy $1.81<M(K \pi)<1.92\gev$
were kept.
Moreover, the $\dz$ candidates were required to have
either
$p_T(\dz)<3.8\gev$ or $|\eta(\dz)|>1.6$.
Any additional track, with $p_T>0.2\gev$
and a charge opposite to that of the kaon track,
was assigned the pion mass and combined with the $\dz$ candidate
to form a $\dsp$ candidate
with invariant mass $M(K \pi \pi_s)$.
Here again the $p_T$ cut was raised to $0.25\gev$ for the data subsample
for which the low-momentum track reconstruction
efficiency was smaller.

\begin{figure}[hbtp]
\epsfysize=18cm
\vspace*{-2.0cm}
\centerline{\epsffile{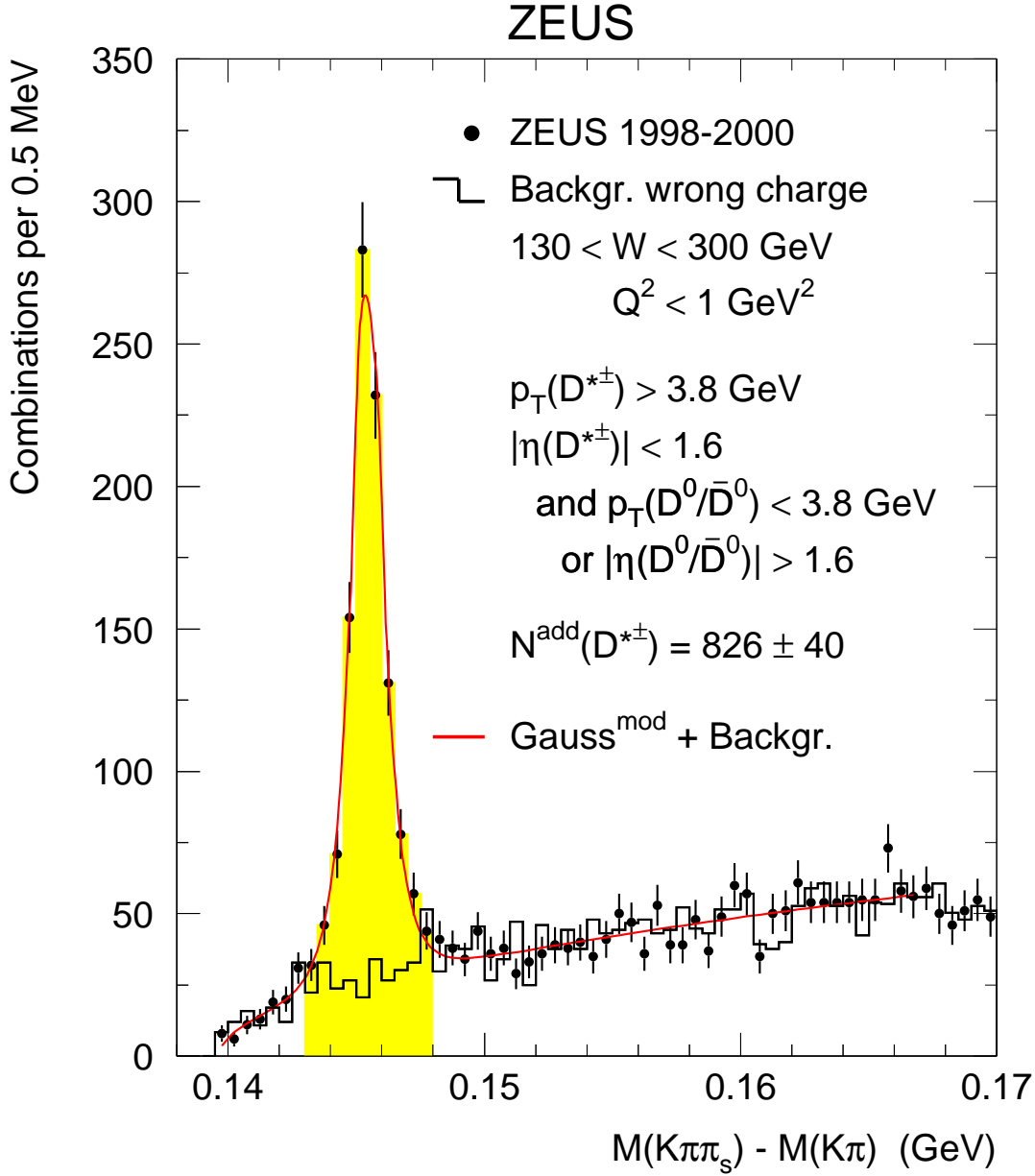}}
\caption{
The distribution of the mass difference,
$\Delta M=M(K \pi \pi_s)-M(K \pi)$, for the ``additional''
$\dspm$  candidates (dots).
The histogram
shows the $\Delta M$ distribution for wrong-charge combinations.
The shaded band shows the signal range in which the wrong-charge
background subtraction was performed.
}
\label{fig:ds}
\end{figure}

Figure~\ref{fig:ds} shows
the $\Delta M$ distribution
for the
$\dsp$ candidates after all cuts.
A clear signal is seen at the nominal value of
$M(\dsp)-M(\dz)$.
The combinatorial background was estimated from
the mass-difference distribution for wrong-charge combinations,
in which both tracks forming the $\dz$ candidate have the same charge
and the third track has the opposite charge.
The same tracks from a wrong-charge combination
can produce two $D^0$ candidates 
due to an ambiguity in the kaon and pion mass assignment to tracks
with the same charge.
To exclude double counting, the multiple combinations of the same tracks
which passed all cuts,
including the $M(K \pi)$ requirement,
were included
with a weight $1/2$.

The number of reconstructed additional $D^{*+}$ mesons
was determined by subtracting
the wrong-charge $\Delta M$ distribution after normalising it
to the distribution of $\dsp$ candidates
with the appropriate charges in the range
$\,0.15<\Delta M<0.17\gev$.
The subtraction, performed in the signal range
$0.143<\Delta M<0.148\gev$, yielded
$N^{\rm add}(\dsp)=826\pm40$.

The $\Delta M$ distribution was also fitted
to a sum of the modified
Gaussian function
(Eq.~(\ref{eq:gausmod}))
describing the signal
and a threshold function
describing the non-resonant background.
The threshold function had a form
$A\cdot (\Delta M -m_\pi)^B$,
where $m_\pi$ is the pion mass~\cite{pl:b592:1}
and $A$ and $B$ were free parameters.
The results obtained using the fit instead of the subtraction
procedure were used to estimate the systematic uncertainty
of the
signal extraction procedure.

\subsection{Reconstruction of $\dc$ mesons}
\label{sec-recdc}

The $\dc$ mesons were reconstructed
from the decay
$D^+ \rightarrow K^-\pi^+\pi^+$.
In each event, two tracks with the same charges and
$p_T>0.5\gev$ and
a third track with opposite charge
and $p_T>0.7\gev$
were combined to form $\dc$ candidates.
The pion masses were assigned to the two tracks
with the same charges and the kaon mass was assigned to the third track,
after which the candidate invariant mass, $M(K\pi\pi)$, was calculated.
To suppress the combinatorial background,
a cut of
$\cos\theta^*(K)>-0.75$ was imposed, where $\theta^*(K)$
is the angle between the kaon in the $K\pi\pi$ rest frame and the $K\pi\pi$
line of flight in the laboratory frame.
To suppress background from $\dsp$ decays, combinations with
$M(K\pi\pi)-M(K\pi)<0.15\gev$ were removed.
The background from
$\dssp \rightarrow \phi\pi^+$ with $\phi \rightarrow K^+K^-$
was suppressed by requiring that the invariant mass of any two
$\dc$ candidate tracks
with opposite charges
was not within $\pm8\mev$ of the $\phi$ mass~\cite{pl:b481:213}
when the kaon mass was assigned to both tracks.

\begin{figure}[hbtp]
\epsfysize=18cm
\vspace*{-2.0cm}
\centerline{\epsffile{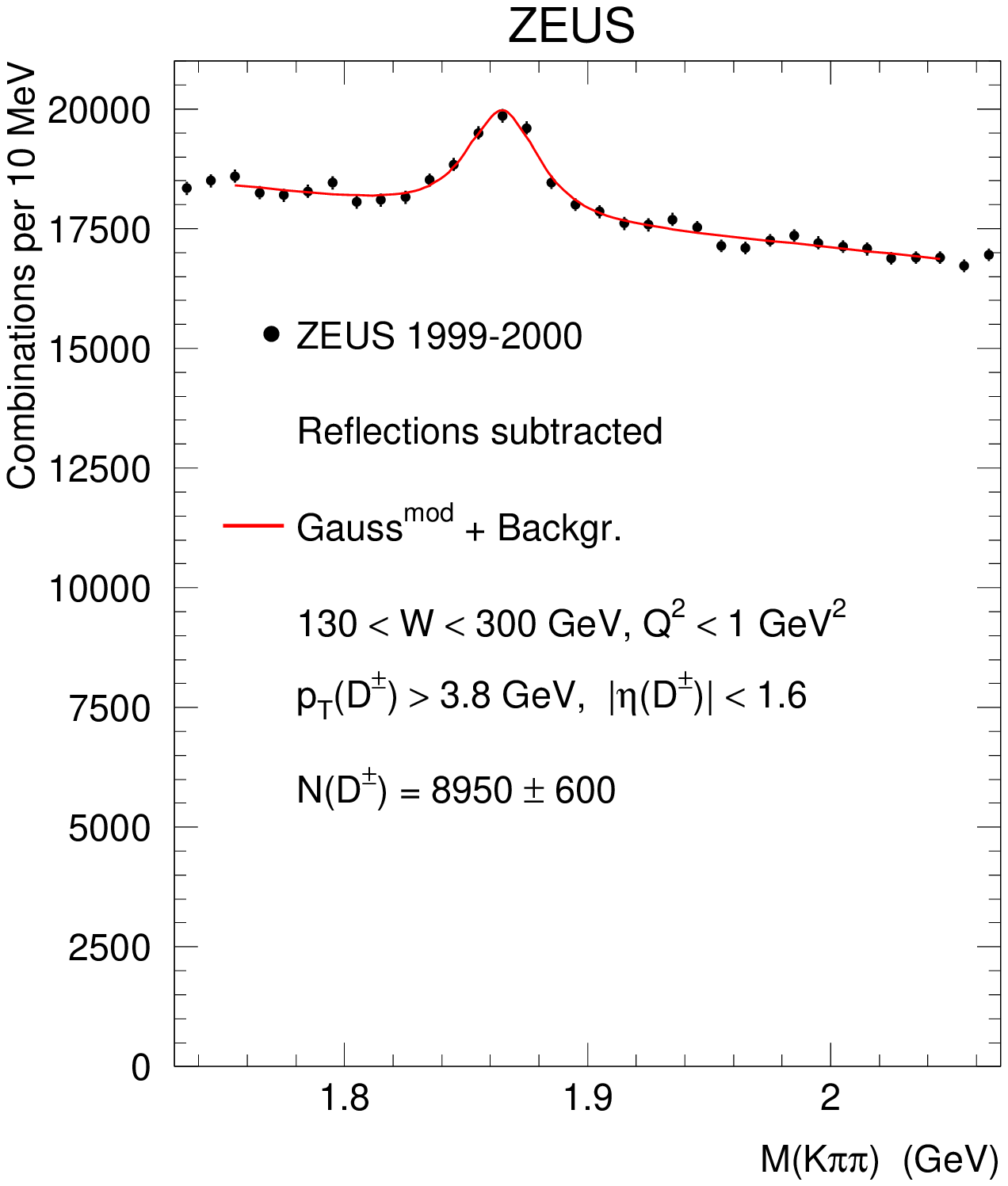}}
\caption{
The $M(K\pi\pi)$ distribution for
the $\dcpm$ candidates (dots).
The solid curve represents a fit to the sum of a modified Gaussian
function and a linear background function.
}
\label{fig:dc}
\end{figure}

Figure~\ref{fig:dc} shows the $M(K\pi\pi)$ distribution for
the $\dc$ candidates after all cuts.
Reflections from $\dssp$ and $\lcp$
decays to three charged particles
were subtracted using the simulated reflection shapes
normalised to the measured $\dssp$ and $\lcp$ production rates.
A clear signal is seen at the nominal value of $D^+$ mass.
The mass distribution was fitted to a sum of a modified
Gaussian function
(Eq.~(\ref{eq:gausmod}))
describing the signal
and a linear function describing the non-resonant background.
The number of reconstructed $\dc$ mesons yielded by the fit was
$N(\dc)=8950\pm600$.

\subsection{Reconstruction of $\dssp$ mesons}
\label{sec-recdss}

The $\dssp$ mesons were reconstructed
from the decay
$\dssp \rightarrow \phi\pi^+$ with $\phi \rightarrow K^+K^-$.
In each event, tracks with opposite charges and
$p_T>0.7\gev$
were assigned the kaon mass and
combined in pairs to form $\phi$ candidates.
The $\phi$ candidate was kept if its invariant mass, $M(KK)$,
was within $\pm8\mev$ of the $\phi$ mass~\cite{pl:b481:213}.
Any additional track  with $p_T>0.5\gev$
was assigned the pion mass and combined with the $\phi$ candidate
to form a $\dssp$ candidate
with invariant mass $M(K K \pi)$.
To suppress the combinatorial background,
the following requirements were applied:
\begin{itemize}
\item{
$\cos\theta^*(\pi)<0.85$, where $\theta^*(\pi)$
is the angle between the pion in the $K K\pi$ rest frame and the $K K\pi$
line of flight in the laboratory frame;
}
\item{
$|\cos^3\theta^{\prime}(K)|>0.1$, where $\theta^{\prime}(K)$
is the angle between one of the kaons and the pion in
the $KK$ rest frame.
The decay of the pseudoscalar $\dssp$ meson to the $\phi$ (vector)
plus $\pi^+$ (pseudoscalar) final state results in an alignment of
the spin of the $\phi$ meson with respect to the direction of
motion of the $\phi$ relative to $\dssp$.
Consequently, the distribution of $\cos\theta^{\prime}(K)$
follows a $\cos^2\theta^{\prime}(K)$ shape, implying a flat
distribution for $\cos^3\theta^{\prime}(K)$.
In contrast, the $\cos\theta^{\prime}(K)$ distribution of
the combinatorial background is flat and its $\cos^3\theta^{\prime}(K)$
distribution peaks at zero. The cut suppressed the background significantly
while reducing the signal by $10\%$.
}
\end{itemize}

\begin{figure}[hbtp]
\epsfysize=18cm
\vspace*{-2.0cm}
\centerline{\epsffile{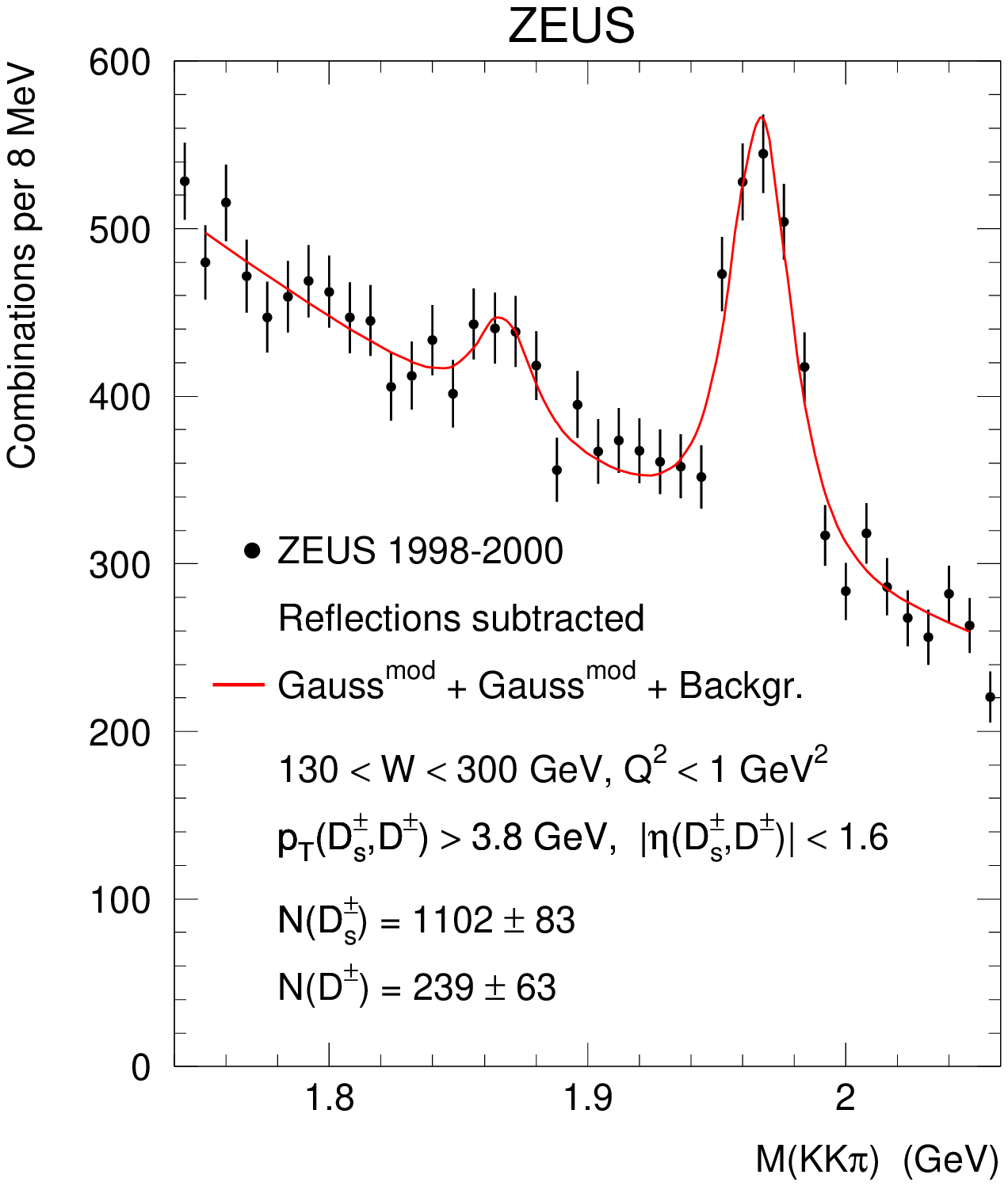}}
\caption{
The $M(K K\pi)$ distribution for
the $\dsspm$ candidates (dots).
The solid curve represents a fit to the sum of
two modified Gaussian functions and an exponential
background function.
}
\label{fig:dss}
\end{figure}

Figure~\ref{fig:dss} shows the $M(K K\pi)$ distribution for
the $\dssp$ candidates after all cuts.
Reflections from $\dc$ and $\lcp$
decays to three
charged particles
were subtracted using the simulated reflection shapes
normalised to the measured $\dc$ and $\lcp$ production rates.
A clear signal is seen at the nominal $\dssp$ mass.
There is also a smaller signal around the nominal $\dc$ mass
as expected from the decay
$\dc \rightarrow \phi\pi^+$ with $\phi \rightarrow K^+K^-$.
The mass distribution was fitted to a sum of two modified
Gaussian functions
(Eq.~(\ref{eq:gausmod}))
describing the signals
and an exponential function describing the non-resonant background.
To reduce the number of free parameters,
the width of the $D^+$ signal was constrained to $8/9$ of the $D^+_s$
signal width; the constraint was verified by MC studies.
The number of reconstructed $\dssp$
mesons yielded by the fit was
$N(\dssp)=1102\pm83$
\footnote{
The number of $D^+$ mesons, $239\pm63$, was not used further in the analysis.}.

\subsection{Reconstruction of $\lcp$ baryons}
\label{sec-reclc}

The $\lcp$ baryons were reconstructed
from the decay
$\lcp \rightarrow K^-p\pi^+$.
In each event, two same-charge tracks
and a third track with opposite charge
were combined to form $\lcp$ candidates.
Due to the large difference between the proton and pion masses, the proton
momentum
is typically larger than that of
the pion.
Therefore, the proton (pion) mass was assigned to those of the two tracks with
the same charges which had larger (smaller) momentum. 
The kaon mass was assigned to the third track and
the candidate invariant mass, $M(K p\pi)$, was calculated.
Only candidates with $p_T(K)>0.75\gev$, $p_T(p)>1.3\gev$
and $p_T(\pi)>0.5\gev$ were kept.
To suppress the combinatorial background,
the following requirements, motivated by MC studies, were applied:
\begin{itemize}
\item{
$\cos\theta^*(K)>-0.9$, where $\theta^*(K)$
is the angle between the kaon in the $K p\pi$ rest frame and the $K p\pi$
line of flight in the laboratory frame;
}
\item{
$\cos\theta^*(p)>-0.25$, where $\theta^*(p)$
is the angle between the proton in the $K p\pi$ rest frame and the $K p\pi$
line of flight in the laboratory frame;
}
\item{
$p^*(\pi)>90\mev$, where $p^*(\pi)$
is the pion momentum in the $K p\pi$ rest frame.
}
\end{itemize}

To suppress the combinatorial background further,
the measured $dE/dx$ values of the three $\lcp$ candidate
tracks were used. The parametrisations of the $dE/dx$ expectation values
and the $\chi^2_1$ probabilities $l_p$, $l_K$ and $l_\pi$ of the proton, kaon
and pion hypotheses, respectively, were obtained in the same way as described
in a previous publication~\cite{epj:c38:29}.
The $l_p$, $l_K$ and $l_\pi$ distributions
for the $\lcp$ candidate tracks
show sharp peaks around
zero and become relatively flat towards one.
To maximise the ratios of the numbers of correctly assigned protons,
kaons and pions to the square roots of the numbers of background
particles, the cuts $l_p>0.15$, $l_K>0.03$ and $l_\pi>0.01$ were
applied. The cuts rejected those ranges where the $l_p$, $l_K$ and $l_\pi$
distributions were at least twice as high as in
the range $0.8-1$.

\begin{figure}[hbtp]
\epsfysize=18cm
\vspace*{-2.0cm}
\centerline{\epsffile{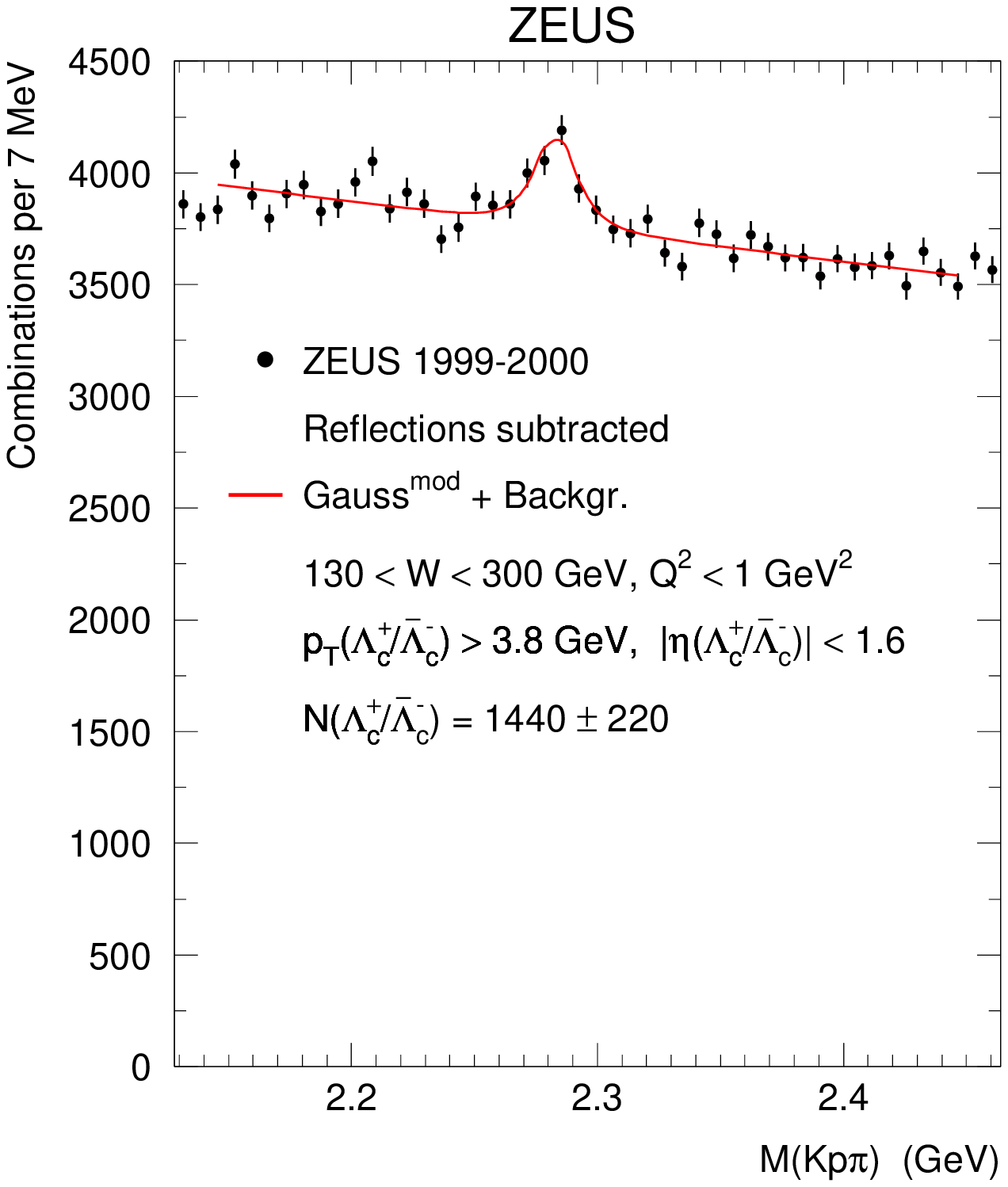}}
\caption{
The $M(K p\pi)$ distribution for
the $\Lambda_c^+ /{\bar \Lambda}_c^{^{-}}$ candidates (dots).
The solid curve represents a fit to the sum of a modified Gaussian
function and a linear background function.
}
\label{fig:lc}
\end{figure}

Figure~\ref{fig:lc} shows the $M(K p\pi)$ distribution for
the $\lcp$ candidates after all cuts.
Reflections from $\dc$ and $\dssp$
decays to three
charged particles were subtracted using the simulated reflection shapes
normalised to the measured $\dc$ and $\dssp$ production rates.
A clear signal is seen at the nominal $\lcp$ mass.
The mass distribution was fitted to a sum of a modified
Gaussian function
(Eq.~(\ref{eq:gausmod}))
describing the signal
and a linear function describing the non-resonant background.
The number of reconstructed $\lcp$ baryons yielded by the fit was
$N(\lcp)=1440\pm220$.

\section{Charm-hadron production cross sections}
\label{sec-xsec}

The charm-hadron cross sections were calculated
for the process  $e p\rightarrow e D(\Lambda_c) X$
in the kinematic region $Q^2<1\gev^2$,
$\,\,\,130<W<300\gev$,
$p_T(D,\Lambda_c)>3.8\gev$ and $|\eta(D,\Lambda_c)|<1.6$.
The cross section for a given charm hadron was calculated from
\begin{equation*}
\sigma(D,\Lambda_c)=\frac{N(D,\Lambda_c)}{\acc \cdot \lum \cdot \bran}\,,
\end{equation*}
where $N(D,\Lambda_c)$ is the number of reconstructed charm hadrons,
$\acc$ is the acceptance for this charm hadron,
$\lum$ is the integrated luminosity
and $\bran$ is the branching ratio or the product of the
branching ratios~\cite{pl:b592:1} for the decay channel
used in the reconstruction.
The third uncertainties quoted below for
the measured cross sections and charm fragmentation ratios and
fractions are due to
the branching-ratio uncertainties\footnote{Contributions from uncertainties of
different branching ratios were added in quadrature.}.

The combined {\sc Pythia}+{\sc Rapgap} MC sample
was used to evaluate the nominal acceptances.
Small admixtures to the reconstructed signals from
other decay modes were taken into account
in the acceptance correction procedure.
To correct from $N^{\rm tag}(\dz)$ ($N^{\rm untag}(\dz)$) to
the production cross sections for $\dz$ mesons originating (not originating)
from $\dsp$ decays, small migrations between the two samples
were taken into account.
The $b$-quark relative contributions, predicted by the MC simulation
using branching ratios of $b$-quark decays to the charmed hadrons
measured at LEP~\cite{pl:b388:648,epj:c1:439},
were subtracted from all measured cross sections\footnote{The branching
ratios of the $b$-quark decays were updated using recent
values~\cite{pl:b592:1} of the relevant charm-hadron decay branching ratios.}.
Subtraction of the $b$-quark contribution reduced the measured cross
sections by $3-7\%$ and
changed the measured charm fragmentation ratios and fractions
by less than $4\%$.

Using the reconstructed signals (see Section~\ref{sec-rec})
the following cross sections
for the sum of each charm hadron and its antiparticle
were calculated.
The systematic uncertainties are discussed in Section~\ref{sec-syst}:

\begin{itemize}
\item{
the production cross section for $\dz$ mesons
not originating from the $\dsp \rightarrow \dz \pi^+_s$ decays:
$$\sigma^{\rm untag}(\dz)=8.49\pm0.44({\rm stat.})^{+0.47}_{-0.48}({\rm syst.})^{+0.20}_{-0.19}({\rm br.})\,{\rm nb};$$
}
\item{
the production cross section for $\dz$ mesons originating
from the $\dsp \rightarrow \dz \pi^+_s$ decays:
$$\sigma^{\rm tag}(\dz)=2.65\pm0.08({\rm stat.})^{+0.11}_{-0.10}({\rm syst.})\pm0.06({\rm br.})\,{\rm nb}.$$
The ratio $\sigma^{\rm tag}(\dz)/\br$ gives
the $\dsp$
cross section, $\sigma(\dsp)$,
corresponding to $\dz$ production in the kinematic range
$p_T(\dz)>3.8\gev$ and $|\eta(\dz)|<1.6$
for the $\dsp \rightarrow \dz \pi^+_s$ decay.
Here $\br=0.677\pm0.005$~\cite{pl:b592:1}
is the branching ratio of the $\dsp \rightarrow \dz \pi^+_s$ decay;
}
\item{
the production cross section for additional $\dsp$ mesons:
$$\sigma^{\rm add}(\dsp)=1.05\pm0.07({\rm stat.})^{+0.09}_{-0.04}({\rm syst.})\pm0.03({\rm br.})\,{\rm nb}.$$
The sum $\sigma^{\rm add}(\dsp)+\sigma^{\rm tag}(\dz)/\br$
gives the production cross section for $\dsp$ mesons
in the kinematic range
$p_T(\dsp)>3.8\gev$ and $|\eta(\dsp)|<1.6$:
$$\sigma^{\rm kin}(\dsp)=4.97\pm0.14({\rm stat.})^{+0.23}_{-0.18}({\rm syst.})^{+0.13}_{-0.12}({\rm br.})\,{\rm nb};$$
}
\item{
the production cross section for $\dc$ mesons:
$$\sigma(\dc)=5.07\pm0.36({\rm stat.})^{+0.44}_{-0.23}({\rm syst.})^{+0.34}_{-0.30}({\rm br.})\,{\rm nb};$$
}
\item{
the production cross section for $\dssp$ mesons:
$$\sigma(\dssp)=2.37\pm0.20({\rm stat.})\pm 0.20({\rm syst.})^{+0.72}_{-0.45}({\rm br.})\,{\rm nb};$$
}
\item{
the production cross section for $\lcp$ baryons:
$$\sigma(\lcp)=3.59\pm0.66({\rm stat.})^{+0.54}_{-0.66}({\rm syst.})^{+1.15}_{-0.70}({\rm br.})\,{\rm nb}.$$
}
\end{itemize}

\section{Charm fragmentation ratios and fractions}
\label{sec-ratio}

\subsection{Ratio of neutral to charged $D$-meson production rates}
\label{sec-rud}

Neglecting influences from decays of heavier excited $D$ mesons,
the ratio of neutral to charged $D$-meson production rates
is given by the ratio of the sum of $D^{*0}$ and direct $\dz$
production cross sections to the sum
of $\dsp$ and direct $\dc$
production cross sections:
$$R_{u/d} =
\frac{\sigma(D^{*0})+\sigma^{\rm dir}(D^0)}
{\sigma(D^{*+})+\sigma^{\rm dir}(\dc)},$$
where $\sigma^{\rm dir}(D^0)$ and $\sigma^{\rm dir}(\dc)$ are those parts of the $\dz$ and $\dc$
inclusive cross sections which
do not originate from $D^{*0}$ and $\dsp$ decays.
Since all $D^{*0}$ decays produce a $\dz$ meson~\cite{pl:b592:1},
the sum of $\sigma(D^{*0})$ and $\sigma^{\rm dir}(D^0)$ is the production
cross section for $D^0$ mesons not originating from $\dsp$ decays:
\begin{equation}
\sigma(D^{*0})+\sigma^{\rm dir}(D^0)=\sigma^{\rm untag}(\dz).
\label{eq:dzdir}
\end{equation}
Subtracting from $\sigma(\dc)$ the contribution from $\dsp$ decays gives
\begin{equation}
\sigma^{\rm dir}(\dc)=\sigma(\dc)-\sigma(D^{*+})\cdot (1-\br).
\label{eq:dcdir}
\end{equation}
Thus, the ratio of neutral and charged $D$-meson production rates
can be calculated as
$$R_{u/d}
= \frac{\sigma^{\rm untag}(D^0)}{\sigma(D^{+})+\sigma(D^{*+})\cdot \br}
=\frac{\sigma^{\rm untag}(D^0)}{\sigma(D^{+})+\sigma^{\rm tag}(D^0)}.$$
Using the measured cross sections, the ratio of neutral to charged $D$-meson production rates,
obtained for the kinematic region $Q^2<1\gev^2$,
$130<W<300\gev$,
$p_T(D)>3.8\gev$ and $|\eta(D)|<1.6$,
is
$$R_{u/d}= 1.100 \pm 0.078 \,({\rm stat.}) ^{+0.038}_{-0.061}\,({\rm syst.})
^{+0.047}_{-0.049}\,({\rm br.}).$$

The measured $R_{u/d}$ value agrees with unity, i.e.$\:$it is
consistent with isospin
invariance, which implies that $u$ and $d$ quarks are produced equally
in charm fragmentation.

Table~\ref{tab:rud} compares the measurement with the values obtained
in DIS~\cite{epj:c38:447} and
in $e^+e^-$ annihilations. The latter value
was calculated as
$$R_{u/d}=\frac{\fcdz-\fcds\cdot\br}{\fcdc+\fcds\cdot\br}$$
using fragmentation fractions
compiled previously~\cite{hep-ex-9912064p}
and updated with the recent branching ratio values~\cite{pl:b592:1}.
All measurements agree with unity within experimental uncertainties.
The branching ratio uncertainties
of all measurements
are highly correlated.

\subsection{Equivalent phase-space treatment}
\label{sec-equ}

In the subtraction of the $\dsp$ contribution to $\dc$ production
in Eq.~(\ref{eq:dcdir}),
the cross-section $\sigma(\dsp)$,
corresponding to $\dz$ production in the kinematic range
$p_T(\dz)>3.8\gev$ and $|\eta(\dz)|<1.6$ for
the $\dsp \rightarrow \dz \pi^+_s$ decay,
was used.
Replacing $\sigma(D^{*+})$ with $\sigma^{\rm tag}(\dz)/\br$ gives
\begin{equation*}
\sigma^{\rm dir}(\dc)=\sigma(\dc)-\sigma^{\rm tag}(\dz)\cdot (1-\br)/\br\,.
\end{equation*}
To compare direct $\dc$ and $\dsp$ production,
the cross section $\sigma^{\rm kin}(\dsp)$ for
$p_T(\dsp)>3.8\gev$ and $|\eta(\dsp)|<1.6$
is used in Section~\ref{sec-pv}.
To compare the inclusive $\dc$ and $\dz$ cross sections with each other
and with the inclusive $\dsp$ cross section
it is necessary
to take into account that
only a fraction of the parent  $D^*$
momentum is transfered
to the daughter $D$ meson.
For such comparisons,
the ``equivalent'' $\dc$ and $\dz$ cross sections were defined
as the sums
of their direct cross sections and contributions from $D^*$ decays
calculated using $\sigma^{\rm kin}(\dsp)$ and  $\sigma^{\rm kin}(D^{*0})$:
\begin{eqnarray*}
\sigma^{\rm eq}(\dc)&=&\sigma^{\rm dir}(\dc) + \sigma^{\rm kin}(\dsp) \cdot (1-\br),\\
\sigma^{\rm eq}(\dz)&=&\sigma^{\rm dir}(\dz) + \sigma^{\rm kin}(\dsp) \cdot \br + \sigma^{\rm kin}(D^{*0}),
\end{eqnarray*}
where $\sigma^{\rm kin}(D^{*0})$ is the inclusive $D^{*0}$ cross section
for $p_T(D^{*0})>3.8\gev$ and $|\eta(D^{*0})|<1.6$.
This cross section can be written as the sum
$\sigma(D^{*0})+\sigma^{\rm add}(D^{*0})$,
where $\sigma(D^{*0})$ is
the part contributing to the $\dz$ production in the nominal kinematic
range (as in Eq.~(\ref{eq:dzdir})) and
$\sigma^{\rm add}(D^{*0})$ is
the production
cross section for ``additional'' $D^{*0}$ mesons
producing $\dz$ mesons outside of that kinematic range.
The latter cross section was calculated using $\sigma^{\rm add}(D^{*+})$
and the expression for $R_{u/d}$:
$$\sigma^{\rm add}(D^{*0})
= \sigma^{\rm add}(D^{*+}) \cdot R_{u/d}
= \sigma^{\rm add}(D^{*+}) \cdot \frac{\sigma^{\rm untag}(D^0)}{\sigma(D^{+})+\sigma^{\rm tag}(D^0)}.$$

Using Eqs.~(\ref{eq:dzdir}) and~(\ref{eq:dcdir}) for
$\sigma^{\rm dir}(\dz)$ and $\sigma^{\rm dir}(\dc)$, respectively, and
the expressions for $\sigma^{\rm kin}(D^{*0})$ and $\sigma^{\rm kin}(D^{*+})$
gives
\begin{eqnarray*}
\sigma^{\rm eq}(\dz)&=&\sigma^{\rm untag}(\dz)+\sigma^{\rm tag}(\dz) + \sigma^{\rm add}(D^{*+}) \cdot (R_{u/d}+\br) ,\\
\sigma^{\rm eq}(\dc)&=&\sigma(\dc) + \sigma^{\rm add}(\dsp) \cdot (1-\br).
\end{eqnarray*}
MC studies show that such ``equivalent phase-space treatment''
for the non-strange  $D$ and $D^*$ mesons
minimises differences between the fragmentation ratios and
fractions measured in the accepted $p_T(D,\Lambda_c)$ and $\eta(D,\Lambda_c)$
kinematic region and those in the
full phase space (see Section 7.6).

\subsection{Strangeness-suppression factor}
\label{sec-gs}

The strangeness-suppression factor for charm mesons
is given by the ratio of twice the production rate of charm-strange mesons
to the production rate of non-strange charm mesons.
All $\dsp$ and $D^{*0}$ decays produce either a $\dc$ or
a $D^0$ meson, while
all $D_s^{*+}$ decays produce a $\dssp$ meson~\cite{pl:b592:1}.
Thus, neglecting decays of heavier excited charm-strange mesons
to non-strange charm mesons,
the strangeness-suppression factor
can be calculated as
a ratio of twice the $\dssp$ production cross section to the sum
of $\dz$ and $\dc$ production cross sections.
Using the equivalent $\dz$ and $\dc$ cross sections gives
$$\gamma_s = \frac{2 \, \sigma(D^+_s)}{\sigma^{\rm eq}(D^+) + \sigma^{\rm eq}(D^0)}
= \frac{2 \, \sigma(D^+_s)}{\sigma(D^+) + \sigma^{\rm untag}(D^0) + \sigma^{\rm tag}(D^0) + \sigma^{\rm add}(D^{*+})\cdot (1+R_{u/d})}.$$
Using the measured cross sections,
the strangeness-suppression factor,
obtained for the kinematic region $Q^2<1\gev^2$,
$130<W<300\gev$,
$p_T(D)>3.8\gev$ and $|\eta(D)|<1.6$, is
$$\gamma_s = 0.257 \pm 0.024 \,({\rm stat.}) ^{+0.013}_{-0.016} \,({\rm syst.}) ^{+0.078}_{-0.049} \,({\rm br.}).$$

Thus, charm-strange meson production is suppressed by a factor $\approx3.9$
in charm fragmentation.
In simulations based on the Lund string fragmentation
scheme~\cite{cpc:39:347,*cpc:43:367}, strangeness suppression is
a free parameter
which determines the ratio of probabilities to create $s$ to $u$ and
$d$ quarks during
the fragmentation processes.
In the absence of excited charm-strange meson decays to non-strange
charm mesons, the Lund strangeness-suppression parameter would be
effectively the observable, $\gamma_s$.
In fact, production rates of the excited charm-strange mesons
are poorly known; varying these rates
in wide ranges in the {\sc Pythia}
simulation suggests that
the Lund strangeness-suppression parameter is $10-30\%$ larger than
the observable, $\gamma_s$.

Table~\ref{tab:gs} compares the measurement with the previous
ZEUS 96-97 result, calculated from the ratio of $\dssp$ to $\dsp$
cross sections~\cite{pl:b481:213}, and with the values obtained
for charm production
in DIS~\cite{epj:c38:447} and
in $e^+e^-$ annihilations. The $e^+e^-$ value
was calculated as
$$\gamma_s = \frac{2 \fcdss} {\fcdc + \fcdz}$$
using fragmentation fractions
compiled previously~\cite{hep-ex-9912064p}
and updated with the recent branching ratio values~\cite{pl:b592:1}.
All measurements agree within experimental uncertainties.
The large branching-ratio uncertainties are dominated by the common
uncertainty of the ${\dssp}\rightarrow\phi\pi^{+}$ branching ratio.
This uncertainty can be ignored in the comparison with other
measurements using the same branching ratios.

\subsection{Fraction of charged $D$ mesons produced in a vector state}
\label{sec-pv}

Neglecting influences from decays of heavier excited $D$ mesons,
the fraction of $D$ mesons produced in a vector state
is given by the ratio of vector to (vector+pseudoscalar)
charm meson production cross sections. Only direct parts
of the production cross sections for pseudoscalar charm mesons
should be used.
Using the expressions for $\sigma^{\rm kin}(\dsp)$
and $\sigma^{\rm dir}(\dc)$,
the fraction for charged charm mesons is given by
$$P^d_{\rm v}
= \frac{\sigma^{\rm kin}(D^{*+})}{\sigma^{\rm kin}(D^{*+})+\sigma^{\rm dir}(\dc)}
= \frac{\sigma^{\rm tag}(D^0)/\br + \sigma^{\rm add}(D^{*+})}
{\sigma(D^+) + \sigma^{\rm tag}(D^0) + \sigma^{\rm add}(D^{*+})}.$$
Using the measured cross sections,
the fraction of charged $D$ mesons produced in a vector state,
obtained for the kinematic region $Q^2<1\gev^2$,
$130<W<300\gev$,
$p_T(D)>3.8\gev$ and $|\eta(D)|<1.6$,
is
$$P^d_{\rm v} = 0.566 \pm 0.025 \,({\rm stat.}) ^{+0.007}_{-0.022} \,({\rm syst.})^{+0.022}_{-0.023} \,({\rm br.}).$$

The measured $P^d_{\rm v}$
fraction
is considerably smaller than the naive spin-counting prediction of $0.75$.
The predictions of the thermodynamical approach~\cite{zfp:c69:485}
and the string fragmentation approach~\cite{zfp:c72:39},
which both predict $2/3$ for the fraction, are closer to,
but still above, the measured value. 
The BKL model~\cite{pr:d62:074013,misc:berezhnoy:private}, based on
a tree-level perturbative QCD calculation with
the subsequent hadronisation
of the $(c,\bar{q})$ state, predicts $P^d_{\rm v}\approx 0.6$ for charm
production in $e^+e^-$ annihilations where only fragmentation diagrams
contribute.
For charm photoproduction, where both fragmentation and recombination
diagrams contribute, the BKL prediction is $P^d_{\rm v}\approx 0.66$
in the measured kinematic range.

Table~\ref{tab:pv} compares the measurement with the values obtained
in DIS~\cite{epj:c38:447} and
in $e^+e^-$ annihilations. The latter value
was calculated as
$$P^d_{\rm v}=\frac{\fcds}{\fcdc+\fcds\cdot\br}$$
using fragmentation fractions
compiled previously~\cite{hep-ex-9912064p}
and updated with the recent branching ratio values~\cite{pl:b592:1}.
The measured $P^d_{\rm v}$ value is smaller than, but consistent with,
the previous measurements.
The branching-ratio uncertainties
of all measurements
are highly correlated.

\subsection{Charm fragmentation fractions}
\label{sec-ff}

The fraction of $c$ quarks hadronising as a particular charm hadron,
$f(c\rightarrow D,\Lambda_c)$, is given by the ratio
of the production cross section for the hadron to the sum
of the production cross sections
for all charm 
ground states that decay weakly.
In addition to the measured $\dz$, $\dc$, $\dssp$ and $\lcp$
charm ground states, the production cross sections of the charm-strange baryons
$\Xi^{+}_c$, $\Xi^{0}_c$ and $\Omega^0_c$
should be included in the sum.
The production rates for these baryons are expected
to be much lower than that of the $\lcp$
due to strangeness suppression.
The relative rates for the
charm-strange baryons
which decay weakly
were estimated from the non-charm sector
following the LEP procedure~\cite{zfp:c72:1}.
The measured $\Xi^{-}/\Lambda$
and $\Omega^{-}/\Lambda$ relative rates are
$(6.65\pm 0.28)\%$ and
$(0.42\pm 0.07)\%$, respectively~\cite{pl:b592:1}.
Assuming equal production of $\Xi^{0}$ and $\Xi^{-}$ states
and that a similar suppression is applicable to the charm baryons,
the total rate for the three charm-strange baryons relative to the
$\lcp$ state is expected to be about $14\%$.
Therefore the $\lcp$ production cross section was scaled by the
factor $1.14$ in the sum of the production cross sections.
An error of $\pm0.05$ was assigned to the scale factor when evaluating
systematic uncertainties.

Using the equivalent $\dz$ and $\dc$ cross sections,
the sum
of the production cross sections
for all open-charm ground states (gs) is given by
$$\sigma_{\rm gs} = \sigma^{\rm eq}(D^+) + \sigma^{\rm eq}(D^0) + \sigma(D_s^+)
+ \sigma(\Lambda_c^+)\cdot 1.14,$$
which can be expressed as
$$\sigma_{\rm gs} = \sigma(D^+) + \sigma^{\rm untag}(D^0) +
\sigma^{\rm tag}(D^0) +
\sigma^{\rm add}(D^{*+})\cdot (1+R_{u/d}) + \sigma(D_s^+)
+ \sigma(\Lambda_c^+)\cdot 1.14.$$

For the measured cross sections,
$$\sigma_{\rm gs} = 24.9 \pm 1.0 \,({\rm stat.}) ^{+1.7}_{-1.4} \,({\rm syst.})^{+1.6}_{-1.0} \,({\rm br.})\,{\rm nb}.$$

The fragmentation fractions for the measured charm ground states
are given by
\begin{eqnarray*}
f(c\rightarrow D^+)&=&\sigma^{\rm eq}(D^+)/\sigma_{\rm gs}
=[\sigma(D^+)+\sigma^{\rm add}(D^{*+})\cdot (1-\br)]/\sigma_{\rm gs},\\
f(c\rightarrow D^0)&=&\sigma^{\rm eq}(D^0)/\sigma_{\rm gs}\\
&=&[\sigma^{\rm untag}(D^0)+\sigma^{\rm tag}(D^0)+\sigma^{\rm add}(D^{*+})\cdot (R_{u/d}+\br)]/\sigma_{\rm gs},\\
f(c\rightarrow D_s^+)&=&\sigma(D_s^+)/\sigma_{\rm gs},\\
f(c\rightarrow \Lambda_c^+)&=&\sigma(\Lambda_c^+)/\sigma_{\rm gs}.
\end{eqnarray*}
Using $\sigma^{\rm kin}(D^{*+})$,
the fragmentation fraction
for the $\dsp$ state is given by
$$f(c\rightarrow D^{*+}) 
=\sigma^{\rm kin}(D^{*+})/\sigma_{\rm gs}
= [\sigma^{\rm tag}(D^0)/\br+\sigma^{\rm add}(D^{*+})]/\sigma_{\rm gs}.$$

The open-charm fragmentation fractions,
measured in the kinematic
region $Q^2<1\gev^2$,
$130<W<300\gev$,
$p_T(D,\Lambda_c)>3.8\gev$ and $|\eta(D,\Lambda_c)|<1.6$,
are summarised in
Table~\ref{tab:ff}.
The results
are compared
with the values obtained
in DIS~\cite{epj:c38:447} and
with the combined fragmentation fractions
for charm production in $e^+e^-$ annihilations compiled
previously~\cite{hep-ex-9912064p}
and updated with the recent branching-ratio values~\cite{pl:b592:1}.
The branching-ratio uncertainties
of all measurements are highly correlated.
The measurements are consistent
although the measured $f(c\rightarrow D^{*+})$ is smaller and 
$f(c\rightarrow \Lambda_c^+)$ is larger than those
obtained in  $e^+e^-$ annihilations.
About half of the difference in the $f(c\rightarrow D^{*+})$ values
is due to the difference in the $f(c\rightarrow \Lambda_c^+)$ values.
The measurement may indicate an enhancement of $\lcp$ production
in $ep$ collisions with respect to $e^+e^-$. However, this is unlikely to be
a consequence of
the baryon-number-flow effect~\cite{pl:b446:321,*hep-ph-0006325}
because no significant asymmetry between the $\Lambda_c^+$ and ${\bar \Lambda}_c^{^{-}}$
production rates was observed\footnote
{Separate fits of the $M(K^-p\pi^+)$ and $M(K^+{\bar p}\pi^-)$ distributions
yielded $N(\Lambda_c^+)/N({\bar{\Lambda}^{^{-}}_c})=0.8\pm0.2$.}.

\subsection{Discussion of extrapolation effects}
\label{sec-ef}

The charm fragmentation ratios and fractions were measured
in the region
$p_T(D,\Lambda_c)>3.8\gev$ and $|\eta(D,\Lambda_c)|<1.6$.
To minimise differences between the 
values measured in the accepted $p_T(D,\Lambda_c)$ and $\eta(D,\Lambda_c)$
kinematic region and those in the
full phase space, the equivalent phase-space treatment
for the non-strange $D$ and $D^*$ mesons was used (see Section~\ref{sec-equ}).

Table~\ref{tab:extrap} shows estimates of extrapolation factors
correcting the values measured in the accepted
$p_T(D,\Lambda_c)$ and $\eta(D,\Lambda_c)$ region to the
full phase space.
The extrapolation factors were determined using three 
different fragmentation schemes:
the Peterson parameterisation~\cite{pr:d27:105} of the charm fragmentation
function as implemented in {\sc Pythia}, 
the Bowler modification~\cite{zfp:c11:169}
of the LUND symmetric fragmentation function~\cite{zfp:c20:317}
as implemented in {\sc Pythia}
and the cluster model~\cite{np:b238:492} as implemented in {\sc Herwig}.
The quoted uncertainties were obtained by varying relevant parameters
in the {\sc Pythia} and {\sc Herwig} MC generators.
The extrapolation factors obtained are generally close to unity.
The only exceptions are the factors
given by the cluster model
for $\fclc$
and, to a lesser extent, for $\gamma_s$ and $\fcdss$.

This MC study suggests that the measured charm fragmentation
ratios and fractions are close to those in the full
$p_T(D,\Lambda_c)$ and $\eta(D,\Lambda_c)$ phase space.

\section{Systematic uncertainties}
\label{sec-syst}

The systematic uncertainties of the measured cross sections
and fragmentation ratios and fractions
were determined
by changing the analysis procedure
and repeating
all calculations.
The following groups of the systematic uncertainty sources
were considered:
\begin{itemize}
\item{$\{\delta_1\}$
the model dependence of the acceptance corrections was estimated using
the {\sc Herwig} MC sample, varying
the $p_T(D,\Lambda_c)$ and
$\eta(D,\Lambda_c)$ distributions of the
reference MC sample
and by changing the MC fraction of charged $D$ mesons produced in a vector
state from 0.6 to $0.5$ or $0.7$;
}
\item{$\{\delta_2\}$
the uncertainty of the beauty subtraction was determined by
varying the
$b$-quark cross section by a factor of two
in the reference MC sample and by varying
the branching ratios of
$b$-quarks to
charm hadrons by their uncertainties~\cite{pl:b388:648,epj:c1:439};
}
\item{$\{\delta_3\}$
the uncertainty of the tracking simulation was obtained
by varying all momenta by $\pm 0.3\%$
(magnetic field uncertainty),
varying the track-loss probabilities by $\pm 20\%$ of their values
and by changing the track momentum and angular resolutions
by $^{+20}_{-10}\%$ of their values.
The asymmetric resolution variations were used since the MC signals
typically had somewhat narrower widths than observed in the data;
}
\item{$\{\delta_4\}$
the uncertainty of the CAL simulation was determined by
varying the CAL energy scale by $\pm 2\%$,
by changing the CAL energy resolution by $\pm 20\%$ of its value
and by varying
the first-level trigger CAL efficiencies;
}
\item{$\{\delta_5\}$
the uncertainties related to the signal extraction procedures were
obtained as follows:}
\begin{itemize}
\item{
for the $\dz$ signals with and without $\Delta M$ tag:
the background parametrisation and the range used for the signal fits
were varied;
}
\item{
for the additional $\dsp$ signal:
the range used for the background normalisation
was varied or
the fit was used instead of the subtraction procedure;
}
\item{
for the $\dc$, $\dssp$ and $\lcp$ signals:
the background parametrisations,
ranges used
for the signal fits
and amounts of the mutual reflections were varied.
In addition, in the $\dssp$ signal-extraction procedure,
the constraint used for the $\dc\rightarrow KK\pi$ signal width was varied.
In the $\lcp$ signal extraction procedure,
an uncertainty in the $dE/dx$ simulation was estimated by changing the $dE/dx$
cut values in the MC and checking the effects with respect to changes
expected from the $\chi^2_1$ distribution.
}
\end{itemize}
\item{$\{\delta_6\}$
the uncertainties of the luminosities of the $e^-p$ ($\pm1.8\%$) and $e^+p$
($\pm2.25\%$) data samples were included taking into account their
correlations;
}
\item{$\{\delta_7\}$
the uncertainty in the rate
of the charm-strange baryons
(see Section~\ref{sec-ff}).
}
\end{itemize}

Contributions from
the different systematic uncertainties were calculated and added
in quadrature separately for positive and negative variations.
The total and $\delta_1$-$\delta_7$ systematic uncertainties
for the charm-hadron
cross sections and charm fragmentation ratios and fractions
are summarised in
Table~\ref{tab:syst}.
Correlated systematic uncertainties largely cancelled
in the calculation of
the fragmentation ratios and fractions.

To check the hadron-mass effects on the measured charm fragmentation
ratios and fractions, the analysis was repeated
using the charm-hadron transverse energy instead of the transverse momentum
in the definition of the kinematic range of the measurement;
the results obtained agreed with the reported values within statistical
errors.
The charm fragmentation ratios and fractions were also calculated separately
for two $W$ sub-ranges; no significant variations were observed.

\section{Summary}
\label{sec-conc}

The production of the charm hadrons
$\dsp$, $\dz$, $\dc$, $\dssp$ and $\lcp$
has been measured with the ZEUS detector
in the kinematic range
$p_T(D,\Lambda_c)>3.8\gev$, $|\eta(D,\Lambda_c)|<1.6$,
$130<W<300\gev$ and $Q^2<1\gev^2$.
The cross sections have
been used to determine
the charm fragmentation ratios and fractions
with comparable precision to the $e^+e^-$ results.

The ratio of neutral to charged $D$-meson production rates
is
$$R_{u/d}= 1.100 \pm 0.078 \,({\rm stat.}) ^{+0.038}_{-0.061}\,({\rm syst.})
^{+0.047}_{-0.049}\,({\rm br.}).$$
The measured $R_{u/d}$ value agrees with unity, i.e.$\:$it is
consistent with isospin invariance,
which implies that $u$ and $d$ quarks are produced equally
in charm fragmentation.

The strangeness-suppression factor is
$$\gamma_s = 0.257 \pm 0.024 \,({\rm stat.}) ^{+0.013}_{-0.016} \,({\rm syst.}) ^{+0.078}_{-0.049`} \,({\rm br.}).$$
Thus, $D_s$-meson production is suppressed by a factor $\approx3.9$
in charm fragmentation.

The fraction of charged $D$ mesons produced in a vector state is
$$P^d_{\rm v} = 0.566 \pm 0.025 \,({\rm stat.}) ^{+0.007}_{-0.022} \,({\rm syst.})^{+0.022}_{-0.023} \,({\rm br.}).$$
The measured fraction
is considerably smaller than the naive spin-counting prediction of $0.75$.
The predictions of the thermodynamical approach~\cite{zfp:c69:485}
and the string fragmentation approach~\cite{zfp:c72:39},
which both predict $2/3$ for the fraction,
and the BKL model~\cite{pr:d62:074013,misc:berezhnoy:private}
prediction ($\approx 0.66$)
are closer to,
but still above, the measured value. 

The measured $R_{u/d}$ and $\gamma_s$ values agree
with those obtained in DIS~\cite{epj:c38:447} and in $e^+e^-$ annihilations.
The $e^+e^-$ values
were calculated
using fragmentation fractions
compiled previously~\cite{hep-ex-9912064p}
and updated with the recent branching ratio values~\cite{pl:b592:1}.
The measured $P^d_{\rm v}$ value is smaller than, but consistent with,
the previous
measurements.

The fractions of $c$ quarks hadronising as
$\dsp$, $\dz$, $\dc$, $\dssp$ and $\lcp$
hadrons have been calculated in the accepted kinematic range.
The measured open-charm fragmentation fractions are
consistent
with previous results 
although the measured $f(c\rightarrow D^{*+})$ is smaller and 
$f(c\rightarrow \Lambda_c^+)$ is larger than those
obtained in  $e^+e^-$ annihilations.
About half of the difference in the $f(c\rightarrow D^{*+})$ values
is due to the difference in the $f(c\rightarrow \Lambda_c^+)$ values.

These measurements generally support the hypothesis that fragmentation
proceeds independently of the hard sub-process.

\section{Acknowledgements}
\label{sec-ackn}

We thank the DESY Directorate for their strong support and encouragement.
The remarkable achievements of the HERA machine group were essential
for the successful completion of this work and
are greatly appreciated.
The design, construction and installation of the ZEUS detector
has been made possible by the efforts of many people who are
not listed as authors.
We thank A.V.~Berezhnoy and A.K.~Likhoded for providing us with
their predictions.

{
\def\bibname{\Large\bf References}
\def\refname{\Large\bf References}
\pagestyle{plain}
\ifzeusbst
  \bibliographystyle{./BiBTeX/bst/l4z_default}
\fi
\ifzdrftbst
  \bibliographystyle{./BiBTeX/bst/l4z_draft}
\fi
\ifzbstepj
  \bibliographystyle{./BiBTeX/bst/l4z_epj}
\fi
\ifzbstnp
  \bibliographystyle{./BiBTeX/bst/l4z_np}
\fi
\ifzbstpl
  \bibliographystyle{./BiBTeX/bst/l4z_pl}
\fi
{\raggedright
\bibliography{./BiBTeX/user/syn.bib,%
              ./BiBTeX/bib/l4z_articles.bib,%
              ./BiBTeX/bib/l4z_books.bib,%
              ./BiBTeX/bib/l4z_conferences.bib,%
              ./BiBTeX/bib/l4z_h1.bib,%
              ./BiBTeX/bib/l4z_misc.bib,%
              ./BiBTeX/bib/l4z_old.bib,%
              ./BiBTeX/bib/l4z_preprints.bib,%
              ./BiBTeX/bib/l4z_replaced.bib,%
              ./BiBTeX/bib/l4z_temporary.bib,%
              ./BiBTeX/bib/l4z_zeus.bib,%
              ./BiBTeX/user/chadr.bib,%
              ./BiBTeX/user/dstargamma.bib,%
              ./BiBTeX/user/eps497.bib}}
}
\vfill\eject

\begin{table}[hbt]
\begin{center}
\begin{tabular}{|c|c|} \hline
& $R_{u/d}$ \\
\hline
\hline
ZEUS ($\gamma p$)
&
$1.100\pm0.078({\rm stat.})^{+0.038}_{-0.061}({\rm syst.})^{+0.047}_{-0.049}({\rm br.})$
\\
\hline
H1 (DIS)~\cite{epj:c38:447}
& $1.26\pm0.20({\rm stat.})\pm0.11({\rm syst.})\pm0.04({\rm br.\oplus theory})$ \\
\hline
combined $e^+e^-$ data~\cite{hep-ex-9912064p}
& $1.020\pm0.069({\rm stat.\oplus syst.})^{+0.045}_{-0.047}({\rm br.})$ \\
\hline
\end{tabular}
\caption{
The ratio of neutral to charged $D$-meson production rates, $R_{u/d}$.
}
\label{tab:rud}
\end{center}
\end{table}
\vspace*{2.cm}
\begin{table}[hbt]
\begin{center}
\begin{tabular}{|c|c|} \hline
& $\gamma_s$ \\
\hline
\hline
ZEUS ($\gamma p$)
& $0.257 \pm 0.024({\rm stat.})^{+0.013}_{-0.016}({\rm syst.})^{+0.078}_{-0.049}({\rm br.})$ \\
\hline
ZEUS 96-97~\cite{pl:b481:213}
& $0.27\pm0.04({\rm stat.})^{+0.02}_{-0.03}({\rm syst.}) \pm0.07({\rm br.})$ \\
\hline
H1 (DIS)~\cite{epj:c38:447}
& $0.36\pm0.10({\rm stat.})\pm0.01({\rm syst.})\pm0.08({\rm br.\oplus theory})$ \\
\hline
combined $e^+e^-$ data~\cite{hep-ex-9912064p}
& $0.259\pm0.023({\rm stat.\oplus syst.})^{+0.087}_{-0.052}({\rm br.})$ \\
\hline
\end{tabular}
\caption{
The strangeness-suppression factor in charm fragmentation, $\gamma_s$.
}
\label{tab:gs}
\end{center}
\end{table}
\vspace*{2.cm}
\begin{table}[hbt]
\begin{center}
\begin{tabular}{|c|c|} \hline
& $P^d_{\rm v}$ \\
\hline
\hline
ZEUS ($\gamma p$)
& $0.566\pm0.025({\rm stat.})^{+0.007}_{-0.022}({\rm syst.})^{+0.022}_{-0.023}({\rm br.})$ \\
\hline
H1 (DIS)~\cite{epj:c38:447}
& $0.693\pm0.045({\rm stat.})\pm0.004({\rm syst.})\pm0.009({\rm br.\oplus theory})$ \\
\hline
combined $e^+e^-$ data~\cite{hep-ex-9912064p}
& $0.614\pm0.019({\rm stat.\oplus syst.})^{+0.023}_{-0.025}({\rm br.})$ \\
\hline
\end{tabular}
\caption{
The fraction of charged $D$ mesons produced in a vector state,
$P^d_{\rm v}$.
}
\label{tab:pv}
\end{center}
\end{table}
\begin{table}[hbt!]
\begin{center}
\begin{tabular}{|c|c|c|c|} \hline
& ZEUS ($\gamma p$) & Combined & H1 (DIS) \\
& $p_T(D,\Lambda_c)>3.8\,$GeV & $e^+e^-$ data~\cite{hep-ex-9912064p} & \cite{epj:c38:447} \\
& $|\eta(D,\Lambda_c)|<1.6$ &  & \\
\hline
\hline
&\phantom{~~~~~~~~~} stat.\phantom{~~} syst.\phantom{~} br.
&\phantom{~~~~}stat.$\oplus\,$syst.\phantom{~} br.
&\phantom{~~~~~~~~~~}total \\
\hline
$\fcdc$ & $0.217 \pm 0.014\phantom{~}^{+0.013\, +0.014}_{-0.005\, -0.016}$  &
$0.226\phantom{~} \pm 0.010\phantom{~~} ^{+0.016}_{-0.014}$ & $0.203 \pm 0.026$ \\
\hline
$\fcdz$ & $0.523 \pm 0.021\phantom{~}^{+0.018\, +0.022}_{-0.017\, -0.032}$  &
$0.557\phantom{~} \pm 0.023\phantom{~~} ^{+0.014}_{-0.013}$ & $0.560 \pm 0.046$ \\
\hline
$\fcdss$ & $0.095 \pm 0.008\phantom{~}^{+0.005\, +0.026}_{-0.005\, -0.017}$ &
$0.101\phantom{~} \pm 0.009\phantom{~~} ^{+0.034}_{-0.020}$ & $0.151 \pm 0.055$ \\
\hline
$\fclc$ & $0.144 \pm 0.022\phantom{~}^{+0.013\, +0.037}_{-0.022\, -0.025}$ &
$0.076\phantom{~} \pm 0.007\phantom{~~} ^{+0.027}_{-0.016}$ & \\
\hline
\hline
$\fcds$ & $0.200 \pm 0.009\phantom{~}^{+0.008\, +0.008}_{-0.006\, -0.012}$ &
$0.238\phantom{~} \pm 0.007\phantom{~~}^{+0.003}_{-0.003}$ & $0.263 \pm 0.032$ \\
\hline
\end{tabular}
\caption{
The fractions of $c$ quarks hadronising as a particular charm hadron,
$f(c \rightarrow D, \Lambda_c)$.
The fractions are shown for the $D^+$, $D^0$, $D^+_s$
and $\Lambda_c^+$ charm ground states and for the
$D^{*+}$ state.
}
\label{tab:ff}
\end{center}
\end{table}
\begin{table}[hbt]
\begin{center}
\begin{tabular}{|c|c|c|c|} \hline
& Peterson & Bowler & Cluster model \\
& ({\sc Pythia}) & ({\sc Pythia}) & ({\sc Herwig}) \\
\hline
\hline
$R_{u/d}$ & $0.99^{+0.02}_{-0.00}$  &
$0.99^{+0.02}_{-0.00}$ & $1.00^{+0.01}_{-0.00}$ \\
\hline
$\gamma_s$ & $1.04^{+0.04}_{-0.07}$  &
$1.00^{+0.05}_{-0.04}$ & $1.18^{+0.07}_{-0.05}$ \\
\hline
$P^d_{\rm v}$ & $1.00 \pm 0.02$  &
$0.97^{+0.01}_{-0.00}$ & $0.96^{+0.02}_{-0.01}$ \\
\hline
$\fcdc$ & $1.00^{+0.02}_{-0.01}$  &
$1.02 \pm ^{+0.01}_{-0.02}$ & $0.99^{+0.01}_{-0.03}$ \\
\hline
$\fcdz$ & $0.99 \pm 0.01$  &
$0.98 \pm 0.01$ & $0.96^{+0.00}_{-0.02}$ \\
\hline
$\fcdss$ & $1.03^{+0.03}_{-0.06}$  &
$1.00^{+0.04}_{-0.03}$ & $1.15^{+0.06}_{-0.05}$ \\
\hline
$\fclc$ & $1.01^{+0.02}_{-0.05}$ &
$1.08^{+0.03}_{-0.02}$ & $1.46^{+0.03}_{-0.09}$ \\
\hline
$\fcds$ & $1.00^{+0.02}_{-0.03}$ &
$0.96^{+0.00}_{-0.02}$ & $0.93^{+0.01}_{-0.02}$ \\
\hline
\end{tabular}
\caption{
The estimates of extrapolation factors which
correct charm fragmentation ratios and fractions
measured in the accepted $p_T(D,\Lambda_c)$ and $\eta(D,\Lambda_c)$
region to the full phase space.
For further details, see text.
}
\label{tab:extrap}
\end{center}
\end{table}
\begin{table}[hbt]
\begin{center}
\begin{tabular}{|c|c|c|c|c|c|c|c|c|} \hline
& total & $\delta_1$
& $\delta_2$
& $\delta_3$
& $\delta_4$
& $\delta_5$
& $\delta_6$
& $\delta_7$
\\
& $(\%)$ & $(\%)$ 
& $(\%)$
& $(\%)$
& $(\%)$
& $(\%)$
& $(\%)$
& $(\%)$
\\
\hline
\hline
$\sigma^{\rm untag}(\dz)$
&
$^{+5.5}_{-5.6}$
&
$^{+2.8}_{-0.6}$
&
$^{+1.8}_{-3.4}$
&
$^{+1.1}_{-1.4}$
&
$^{+1.3}_{-1.2}$
&
$^{+3.4}_{-3.4}$
&
$^{+2.2}_{-2.2}$
&
\\
\hline
$\sigma^{\rm tag}(\dz)$
&
$^{+4.0}_{-3.7}$
&
$^{+2.5}_{-1.2}$
&
$^{+1.1}_{-2.1}$
&
$^{+1.4}_{-1.3}$
&
$^{+1.4}_{-1.1}$
&
$^{+0.7}_{-0.4}$
&
$^{+2.2}_{-2.2}$
&
\\
\hline
$\sigma^{\rm add}(\dspm)$
&
$^{+8.4}_{-3.6}$
&
$^{+5.8}_{-0.4}$
&
$^{+1.0}_{-1.9}$
&
$^{+3.3}_{-1.6}$
&
$^{+1.8}_{-1.4}$
&
$^{+4.2}_{-0.1}$
&
$^{+2.2}_{-2.2}$
&
\\
\hline
$\sigma^{\rm kin}(\dspm)$
&
$^{+4.6}_{-3.6}$
&
$^{+3.1}_{-1.0}$
&
$^{+1.1}_{-2.0}$
&
$^{+1.6}_{-1.3}$
&
$^{+1.4}_{-1.1}$
&
$^{+1.1}_{-0.3}$
&
$^{+2.2}_{-2.2}$
&
\\
\hline
$\sigma(\dcpm)$
&
$^{+8.7}_{-4.5}$
&
$^{+4.5}_{-0.3}$
&
$^{+1.8}_{-3.3}$
&
$^{+1.0}_{-1.6}$
&
$^{+1.3}_{-1.0}$
&
$^{+6.7}_{-0.6}$
&
$^{+2.3}_{-2.3}$
&
\\
\hline
$\sigma(\dsspm)$
&
$^{+8.3}_{-8.5}$
&
$^{+6.0}_{-0.0}$
&
$^{+3.9}_{-7.0}$
&
$^{+1.4}_{-1.3}$
&
$^{+1.8}_{-1.0}$
&
$^{+2.9}_{-4.0}$
&
$^{+2.2}_{-2.2}$
&
\\
\hline
$\sigma(\lcpm)$
&
$^{+15.1}_{-18.3}$
&
$^{+13.2}_{\,\,-0.6}$
&
$^{+3.1}_{-5.6}$
&
$^{+4.8}_{-1.5}$
&
$^{+2.2}_{-3.0}$
&
$^{\,\,+3.3}_{-16.9}$
&
$^{+2.3}_{-2.3}$
&
\\
\hline
$R_{u/d}$
&
$^{+3.5}_{-5.5}$
&
$^{+0.0}_{-0.9}$
&
$^{+0.4}_{-0.7}$
&
$^{+0.6}_{-0.6}$
&
$^{+0.1}_{-0.3}$
&
$^{+3.4}_{-5.4}$
&
$^{+0.2}_{-0.2}$
&
\\
\hline
$\gamma_s$
&
$^{+5.0}_{-6.3}$
&
$^{+2.4}_{-0.2}$
&
$^{+2.4}_{-4.1}$
&
$^{+0.9}_{-0.8}$
&
$^{+0.8}_{-0.0}$
&
$^{+3.5}_{-4.7}$
&
$^{+0.1}_{-0.1}$
&
\\
\hline
$P^d_{\rm v}$
&
$^{+1.2}_{-3.9}$
&
$^{+0.4}_{-1.0}$
&
$^{+0.8}_{-0.5}$
&
$^{+0.5}_{-0.1}$
&
$^{+0.1}_{-0.1}$
&
$^{+0.6}_{-3.7}$
&
$^{+0.1}_{-0.1}$
&
\\
\hline
$\sigma_{\rm gs}$
&
$^{+6.8}_{-5.7}$
&
$^{+5.3}_{-0.4}$
&
$^{+2.0}_{-3.8}$
&
$^{+1.8}_{-1.0}$
&
$^{+1.2}_{-1.1}$
&
$^{+1.9}_{-3.1}$
&
$^{+2.2}_{-2.2}$
&
$^{+0.7}_{-0.7}$
\\
\hline
$\fcdc$
&
$^{+6.1}_{-2.1}$
&
$^{+0.3}_{-0.8}$
&
$^{+0.7}_{-0.4}$
&
$^{+0.2}_{-1.0}$
&
$^{+0.6}_{-0.3}$
&
$^{+6.0}_{-1.5}$
&
$^{+0.1}_{-0.1}$
&
$^{+0.7}_{-0.7}$
\\
\hline
$\fcdz$
&
$^{+3.4}_{-3.2}$
&
$^{+0.2}_{-2.2}$
&
$^{+0.9}_{-0.5}$
&
$^{+0.3}_{-0.8}$
&
$^{+0.5}_{-0.4}$
&
$^{+3.1}_{-2.1}$
&
$^{+0.1}_{-0.1}$
&
$^{+0.7}_{-0.7}$
\\
\hline
$\fcdss$
&
$^{+4.9}_{-5.4}$
&
$^{+0.8}_{-0.2}$
&
$^{+1.9}_{-3.3}$
&
$^{+0.3}_{-0.8}$
&
$^{+1.2}_{-0.3}$
&
$^{+4.1}_{-4.1}$
&
$^{+0.1}_{-0.1}$
&
$^{+0.7}_{-0.7}$
\\
\hline
$\fclc$
&
$^{\,\,+9.1}_{-15.1}$
&
$^{+7.4}_{-0.3}$
&
$^{+1.2}_{-1.9}$
&
$^{+3.5}_{-0.7}$
&
$^{+1.7}_{-2.9}$
&
$^{\,\,+3.1}_{-14.6}$
&
$^{+0.2}_{-0.2}$
&
$^{+0.7}_{-0.7}$
\\
\hline
$\fcds$
&
$^{+3.9}_{-3.2}$
&
$^{+0.4}_{-2.2}$
&
$^{+1.9}_{-1.0}$
&
$^{+0.6}_{-0.5}$
&
$^{+0.7}_{-0.5}$
&
$^{+3.1}_{-1.9}$
&
$^{+0.1}_{-0.1}$
&
$^{+0.7}_{-0.7}$
\\
\hline
\end{tabular}
\caption{
The total and $\delta_1$-$\delta_7$ (see text)
systematic uncertainties
for the charm hadron
cross sections and charm fragmentation ratios and fractions.
}
\label{tab:syst}
\end{center}
\end{table}

%
%

%
%
\end{document}